\documentclass{jfm}

\pdfoutput=1

\usepackage{upgreek}

\usepackage{amsmath}
\usepackage{amsfonts}

\newcommand{\Stn}{St}
\newcommand{\xx}{\boldsymbol x}
\newcommand{\av}{\boldsymbol a}
\newcommand{\bv}{\boldsymbol b}
\newcommand{\cv}{\boldsymbol c}
\newcommand{\gv}{\boldsymbol g}
\newcommand{\uv}{\boldsymbol u}

\newcommand{\Cmat}{\mathsfbi C}
\newcommand{\Rmat}{\mathsfbi R}
\newcommand{\Smat}{\mathsfbi S}
\newcommand{\Tmat}{\mathsfbi T}
\newcommand{\Lmat}{\mathsfbi L}
\newcommand{\Umat}{\mathsfbi U}
\newcommand{\Dmat}{\mathsfbi D}
\newcommand{\XX}{\mathsfbi X}
\newcommand{\YY}{\mathsfbi Y}

\newcommand{\ii}{{\mathrm{i}}}

\newcommand{\deltat}{\Updelta t}

\newcommand{\Sigmamat}{{\boldsymbol\Upsigma}}
\newcommand{\Omegamat}{{\boldsymbol\Upomega}}

\DeclareMathOperator{\sgn}{sgn}

\begin{document}

\shorttitle{Spectral proper orthogonal decomposition} 
\shortauthor{M. Sieber et al} 

\title{Spectral proper orthogonal decomposition}

\author
 {
  Moritz Sieber
  \corresp{\email{moritz.sieber@tu-berlin.de}},
  K. Oberleithner
  \and 
  C. O. Paschereit
  }

\affiliation
{
Institut f\"ur Str\"omungsmechanik und Technische Akustik, HFI,\\
 M\"uller-Breslau Str. 8, D-10623 Berlin, Germany
}

\maketitle

\begin{abstract}
The identification of coherent structures from experimental or numerical data is an essential task when conducting research in fluid dynamics. 
This typically involves the construction of an empirical mode base that appropriately captures the dominant flow structures. 
The most prominent candidates are the energy-ranked proper orthogonal decomposition (POD) and the frequency ranked Fourier decomposition and dynamic mode decomposition (DMD). 
However, these methods fail when the relevant coherent structures occur at low energies or at multiple frequencies, which is often the case. 
To overcome the deficit of these ``rigid'' approaches, we propose a new method termed Spectral Proper Orthogonal Decomposition (SPOD).
It is based on classical POD and it can be applied to spatially and temporally resolved data. The new method involves an additional temporal constraint that enables a clear separation of phenomena that occur at multiple frequencies and energies. 
SPOD allows for a continuous shifting from the energetically optimal POD to the spectrally pure Fourier decomposition by changing a single parameter.  
In this article, SPOD is motivated from phenomenological considerations of the POD autocorrelation matrix and justified from dynamical system theory. 
The new method is further applied to three sets of PIV measurements of flows from very different engineering problems. 
We consider the flow of a swirl-stabilized combustor, the wake of an airfoil with a Gurney flap, and the flow field of the sweeping jet behind a fluidic oscillator. 
For these examples, the commonly used methods fail to assign the relevant coherent structures to single modes. 
The SPOD, however, achieves a proper separation of spatially and temporally coherent structures, which are either hidden in stochastic turbulent fluctuations or spread over a wide frequency range. 
The SPOD requires only one additional parameter, which can be estimated from the basic time scales of the flow. 
In spite of all these benefits, the algorithmic complexity and computational cost of the SPOD are only marginally greater than those of the snapshot POD.
\end{abstract}
\section{Introduction and motivation}\label{intro_motivation}

\subsection{Contemporary methods for data reduction}\label{intro}
Today's high fidelity computational fluid dynamics (CFD) and high-end experimental data acquisition systems tend to produce vast amounts of data that are getting harder to interpret and overview. 
Methods to analyze such data are numerous and are always developing to stay in line with acquisition and computation systems. 
The most challenging data stem from turbulent flows that feature a huge range of temporal and spatial scales. 
A key challenge in turbulent flow data mining is the distinction of deterministic coherent motion from purely stochastic motion. 
Numerous methods exist that exploit the periodicity or energetic dominance of these coherent structures. 
These methods range from classic Fourier decomposition to dynamic mode decomposition (DMD) and proper orthogonal decomposition (POD). 
The most prominent among them are shortly introduced in the following.

POD has been widely used since its introduction by \citet{Lumley.1970} and \citet{Sirovich.1987}. It was applied in nearly every fluid dynamic field.
Beyond fluid dynamics, this method is also known as singular value decomposition, principal component analysis or Kahunen-Loeve expansion \citep{Berkooz.1993}. 
The basic idea behind this method is to construct an optimal basis that represents most of the data variance with as few basis functions as possible. 
In context of POD the variance is turbulent kinetic energy. 
Therefore, the POD searches for the most energetic modes whereby coherent structures with high energy content are likely to be represented by POD basis functions \citep{Holmes.2012}.

Another classical approach is the linear stochastic estimation introduced by \citet{Adrian.1988}, where the readings of different sensors are related via a linear mapping. 
This is closely related to the extended POD \citep{Boree.2003}, also described in a unified framework (observable inferred decomposition) by \citet{Schlegel.2012}. 
In recent extensions of linear stochastic estimation, the use of time delays between the different sensors and also the use of one sensor at multiple time instances is pursued to separate periodic coherent structures from turbulent fluctuations \citep{Durgesh.2010,Lasagna.2013}. 
This approach was also used to improve the determination of harmonic POD modes from few pressure sensors \citep{Hosseini.2015}. 
These utilisations of data from various time instances are also related to the temporal constraint used for the POD extension proposed in this article.

Targeting the temporal periodicity of the coherent structures, spectral methods like discrete Fourier transform (DFT) and the recently introduced DMD \citep{Rowley.2009,Schmid.2010} come into play. 
These methods commonly span the mode space according to fixed frequencies, which enables the identification of coherent structures within small spectral bandwidths. 
In contrast to the DFT, the DMD also distinguishes modes with respect to their linear amplification. 
The recently introduced extended DMD \citep{Williams.Submitted} tries to overcome the limitations encountered by the (linear) DMD approach when trying to decompose data from nonlinear systems. 
The idea is to use nonlinear functions that create observables of the data, which are {\itshape{exactly}} described by a linear system. 
This approach translocates the problem towards the identification of these nonlinear functions, which can be solved using the ``kernel trick'' that is common in machine learning. 
This paper presents an alternative approach, which extends POD to account for temporal dynamics in addition to energetic optimality.

\subsection{Why yet another method?}\label{motivation}
After this short and incomplete review of data processing methods, one may ask if there is need for another method. 
The answer is probably no, so we take the most used method (POD) and bring it up to date for present research issues. 
The approach pursued here includes a simple yet effective extension to the classical POD, which leads to a more general method comprising POD and also the DFT. 
This approach unifies existing methods, but also offers possibilities beyond these. 
From the authors' experience, the currently available methods often fail when applied to challenging flow data. 
These stem from flows with weak coherent structures where the recorded data have low signal to noise ratios, from flows  with intermittent dynamics, or from flows featuring multi-modal interactions leading to frequency modulations, to name a few.
In such cases, much effort is required to optimize the data processing until satisfactory results are obtained. 
The usual escape route is to focus on a certain spatial region or to apply suitable filters to pick out a certain wavelength or frequency range. 
This involves trial and error or requires prior knowledge of the investigated flow. There is also the danger of cutting off a substantial portion of the data, leading to false interpretations. 
These procedures can be collected under the heading ``identifying symmetries'' as done by \citet{Holmes.2012}. 
The drawback of this approach is that the investigated flow must feature some symmetries and {\itshape they must be known a priori}. 
A recent application shows the huge variety of spatial and temporal filtering together with POD to separate different phenomena into different modes \citep{Bourgeois.2013}, exemplifying the complexity of this approach. 

The usage of spectral methods for highly turbulent flows is even more challenging than POD. 
The variable frequency of single coherent structures and intermittent occurrence of different structures with the same frequency hinders a proper decomposition. 
In terms of the DFT, averaging of spectra from multiple measurements or sensors is essential to get reliable results. 
Analogously, for the DMD, averaging over several events is an option to reject noise \citep{Tu.2014}. 
Nonetheless, the results obtained with DFT and DMD suffer from limiting the temporal dynamics to single frequencies. 
Turbulent flows hardly ever feature discrete frequencies and it is not always valuable to restrict a single mode (flow phenomenon) to a single frequency. 
Coherent structures that feature significant phase jitter or frequency modulation are represented by many modes at similar frequencies. 
In contrast, the POD puts no temporal constraint on the modes. 
This can result in modes that represent flow phenomena occurring at largely different temporal scales. 
Thus, it is often hard to interpret these modes and draw meaningful conclusions from the temporal dynamics.

From our point of view, there is a big gap between the energetically optimal decomposition of POD and the spectrally clean decomposition of DFT or DMD. 
This gap will be bridged with the spectral proper orthogonal decomposition (SPOD) introduced in this article. 
This new method not only places itself somewhere in between these two extrema, but it allows for a continuous shifting from one to the other. 
The main idea is to apply a filter operation to the POD correlation matrix, which will force the POD towards clear temporal dynamic. 
Depending on the filter strength we continuously sweep from classic POD to DFT. 

The remainder of this article is organized as follows: The proposed method is described in detail in \S\ref{sec:SPOD}. The reader is guided from snapshot POD via an in-depth interpretation of the correlation matrix towards the general description of the SPOD. In addition, a method is explained to identify coupled mode pairs describing a single coherent structure, which becomes handy when working with SPOD. In \S\ref{sec:application}, the new method is demonstrated on three different experimental data sets.  The results are compared against POD and DFT to point out the benefits of SPOD. In \S\ref{sec:summary_conclusion} the capabilities of SPOD are summarized, based on the findings from the application to experimental data.

\section{Description and interpretation of the proposed method}\label{sec:SPOD}

\subsection{Classical snapshot POD}\label{snapshot_POD}

To introduce the method and the nomenclature, the snapshot POD approach is described first. We start off with a decomposition of a data set into spatial modes and temporal coefficients:
\begin{align}
U(\xx,t) = \overline{u}(\xx) + u(\xx,t) = \overline{u}(\xx) + \sum_{i=1}^N a_i(t) \Phi_i(\xx).
\label{eqn:POD_decomposition}
\end{align}
Note that only the fluctuating part $u(\xx,t)$ is decomposed. 
It is split into a sum of spatial modes $\Phi_i$ and mode coefficients $a_i$.
A set of $M$ spatial points recorded simultaneously over $N$ time steps is considered. 
To calculate the POD, the correlation matrix of this data set is needed. 
For data obtained from particle image velocimetry (PIV) or CFD, the number of spatial points is usually larger than the number of snapshots. 
The correlation matrix is then calculated between individual snapshots (temporal correlation). 
The alternative approach (spatial correlation) that applies for $M\gg N$, is detailed in appendix \ref{spatial_SPOD}. 
The correlation between two snapshots is calculated from an appropriate inner product $\left<,\right>$, usually defined as the $L^2$ inner product
\begin{align}
\left<u(\xx),v(\xx)\right> = \int\limits_V{u(\xx)v(\xx)\mathrm{d}V},
\end{align}
where $V$ specifies the spatial region or volume over which the correlation is integrated. 
The elements of the correlation matrix $\Rmat$ are given by
\begin{align}
R_{i,j} = \frac{1}{N}\left<u(\xx,t_i),u(\xx,t_j)\right>.
\end{align}
Matrix $\Rmat$ is of size $N \times N$.

The temporal coefficients $\av_i$ and mode energies $\lambda_i$ are obtained from the eigenvectors and eigenvalues of the correlation matrix.
\begin{align}
\Rmat \av_i = \lambda_i \av_i \ ; \quad \lambda_1 \ge \lambda_2 \ge \cdots \ge \lambda_N \ge 0
\end{align}
The subscript $i$ refers to single eigenvalues, which are sorted in descending order. 
Since the $\av_i$ are the eigenvectors of the real symmetric positive-definite matrix $\Rmat$, they are orthogonal. 
Moreover, they are scaled with the energy of the single modes such that
\begin{align}
\frac{1}{N} (\av_i,\av_j)= \lambda_i \delta_{ij},
\end{align}
where $\left(,\right)$ denotes the scalar product. The spatial modes are obtained from the projection of the snapshots onto the temporal coefficients
\begin{align}
\Phi_i(\xx) = \frac{1}{N \lambda_i} \sum_{j=1}^N{a_i(t_j) u(\xx,t_j)}.
\end{align}
These modes are orthonormal by construction
\begin{align}
\left<\Phi_i,\Phi_j\right> = \delta_{ij}.
\end{align}
The formulation so far is perfectly in line with classical snapshot POD, which can also be computed by a singular value decomposition. 
However, since the SPOD requires a manipulation of the correlation matrix we retain the classical form.

\subsection{Properties of the correlation matrix}\label{corr_matrix}

\begin{figure}
	\centering
		\includegraphics{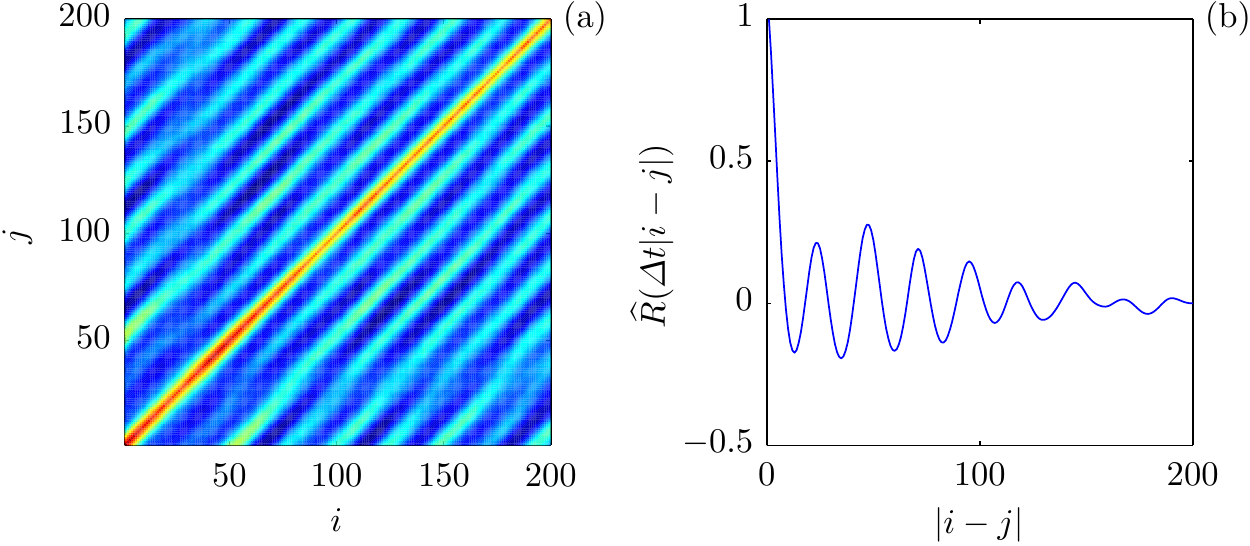}
		\caption{(a) Pseudo-color plot of the correlation matrix elements $R_{i,j}$ and (b) the corresponding correlation coefficient $\widehat{R}$. The displayed data are picked from PIV measurements of a forced turbulent jet. }
	\label{fig:corr_mat}
\end{figure}

The SPOD described in this article is essentially a filter applied to the correlation matrix $\Rmat$.  To offer a better understanding of this approach, the structure of the correlation matrix $\Rmat$
is inspected first. 

Figure \ref{fig:corr_mat}(a) shows the structure of the correlation matrix for the data set of a forced turbulent jet. 
The data were acquired with PIV inside a 2D-plane aligned with the jet axis. 
The considered flow shows strong vortex shedding at the forcing frequency (the acquisition frequency is 25 times the forcing frequency). 
The presence of these periodic patterns in the flow, and their convection within the observed flow field, lead to a diagonal, wave-like structure of the matrix. 
This is closely related to the periodicity of the auto-correlation coefficient. 
In fact, if the individual elements of the correlation matrix $\Rmat$ are summed up along the diagonals, we get the spatially averaged auto-correlation coefficient
\begin{align}
\widehat{R}(\tau) = \frac{\iint{u(\xx,t)u(\xx,t + \tau)}\mathrm{d}\xx\mathrm{d}t}{\iint{u^2(\xx,t)}\mathrm{d}\xx\mathrm{d}t}.
\label{eqn:corr_coeff}
\end{align}
It is depicted in figure \ref{fig:corr_mat}(b), showing the same periodicity as the correlation matrix. 
The auto-correlation coefficient itself represents the spectral content of different timescales and wavelengths and it is directly related to the power spectral density of the underlying data. 
However, it contains no information of the phase of individual frequencies, due to the reference of the signal to itself. 
This is why the elements along the diagonals of $\Rmat$ look so similar, as they represent only relative changes with respect to the time step on the main diagonal. 
Thus, increased similarity along the diagonals of $\Rmat$ is equivalent to an increased similarity of the dynamics of the underlying signal. 
This property will be discussed more deeply in section \ref{SPOD_theory}. 
The obvious consequence from these findings is: If we want to obtain smooth dynamics from the POD, we have to enforce the diagonal similarity of the correlation matrix. 
This is where we step into spectral POD.

\subsection{General description of the SPOD}\label{spectral_POD}

The yet so simple, but radical approach is a filter operating on the correlation matrix $\Rmat$. To augment the diagonal similarity of $\Rmat$ a simple low-pass filter is applied along the diagonals. This results in a filtered correlation matrix $\Smat$ with the elements given as
\begin{align}
S_{i,j} = \sum_{k=-N_f}^{N_f}{g_k R_{i+k,j+k}}. \label{eqn:SPOD_filter}
\end{align}
The filter above is just a symmetric finite impulse response filter with a filter coefficients vector $\gv$ of length $2N_f+1$. The most simple approach would be a box filter, where all coefficients have the same value $g_k = \frac{1}{2N_f+1}$. 
In the examples discussed later, we use a Gaussian filter, which features a smooth response in time and frequency domain. Moreover, we choose a standard deviation as such that the filter gives the same cut-off frequency as a box filter with half the length. In fact, any kind of digital finite impulse response filter can be used here. 

The further procedure of the SPOD is the same as for the classical POD. From the correlation Matrix $\Smat$ the temporal coefficients $\bv_i$ and mode energies $\mu_i$ are obtained from the eigen decomposition
\begin{align}
\Smat \bv_i = \mu_i \bv_i  \ ; \quad \mu_1 \ge \mu_2 \ge \cdots \ge \mu_N \ge 0.
\end{align}
The temporal coefficients are also scaled with the mode energy and they are still orthogonal
\begin{align}
\frac{1}{N}(\bv_i,\bv_j) = \mu_i \delta_{ij}.
\end{align}
The spatial modes are finally obtained from the projection of the snapshots onto the temporal coefficients
\begin{align}
\Psi_i(\xx) = \frac{1}{N \mu_i} \sum_{j=1}^N{b_i(t_j) u(\xx,t_j)},\label{eqn:SPOD_temporal_projection}
\end{align}
where these modes are no longer orthogonal.
This property of the spatial modes is detailed in appendix \ref{SPOD_mode_properties}.
The total energy of the data set is still represented by the decomposition ($\sum{\lambda_i}=\sum{\mu_i}$), but the energy per mode is less for the first modes.
Hence, increasingly plain temporal dynamics are obtained at the expense of spatial orthogonality and a dispersed SPOD spectrum. Nevertheless, the decomposition (as in \eqref{eqn:POD_decomposition})
\begin{align}
U(\xx,t) = \overline{u}(\xx) + \sum_{i=1}^N b_i(t) \Psi_i(\xx),
\end{align}
is still exact if all $N$ SPOD modes are used for the re-composition. 

If the filter size is extended over the entire time-series, the filtered correlation matrix converges to a symmetric Toeplitz matrix. This matrix has the form
\begin{align}
S_{i,j} = \hat{R}\left(\deltat|i-j|\right),
\end{align}
with the diagonals given by the average correlation coefficient \eqref{eqn:corr_coeff}. 
This special matrix is also known as the covariance matrix and its eigenvalues are tracing out the power spectral density of the underlying time series \citep{Wise.1955}. 
This equality is a part of Szeg\"os theorem and it is valid for the limiting case where the number of samples approaches infinity.
To discuss this feature for finite series, the treatment of the start and end of the time series must be clarified. 
At the boundaries of $\Rmat$, the filter operation is not properly defined, since the symmetric filter lacks elements before and after the finite series. 
These elements can either be replaced by zeros or the time-series is assumed to be periodic. 
In the latter case, a box filter of the same size as the number of snapshots delivers a symmetric circulant matrix, with its eigenvalues and eigenvectors given by the Fourier transform of the first row \citep{Gray.2005}.
The DFT and the SPOD produce the same decomposition for this limiting case. 
Hence, the SPOD is able to continuously fade from the energetically optimal POD to a purely spectral DFT. 
What happens in between these two limits is very promising, as will be shown later.

\subsection{The correlation matrix from a dynamic systems point of view}\label{SPOD_theory}

To further consolidate the previous considerations, the correlation matrix is represented in an analytical framework. For the moment, the temporal evolution of the investigated flow is assumed to be locally governed by a linear time invariant model. The term locally refers to a short and finite temporal extension, and it is closely related to the filter size $N_f$. The temporal evolution of this system is given by 
\begin{align}
\frac{\partial \uv(t)}{\partial t} = \Lmat \uv(t),
\end{align}
where the matrix $\Lmat$ is the system matrix describing the entire dynamics and the spatial points of the velocity filed are organized as rows in $\uv$.
Starting from a reference snapshot $\uv_0= \uv(t=0)$, the field at any time step can be calculated by the matrix exponential
\begin{align}
\uv(t) = e^{\Lmat t} \uv_0. \label{eqn:LTI_ansatz_1}
\end{align}
To allow further simplifications, we require the matrix $\Lmat$ to be normal. This allows for the decomposition $\Lmat=\Umat\Dmat\Umat^*$, where $\Umat$ is a unitary matrix (eigenvectors of $\Lmat$), $\Dmat$ is a diagonal matrix (eigenvalues of $\Lmat$) and $^*$ means the conjugate transpose of a matrix (adjoined). Hereafter, equation \eqref{eqn:LTI_ansatz_1} becomes
\begin{align}
\uv(t) = e^{\Umat \Dmat \Umat^* t} \uv_0 = \Umat e^{\Dmat t} \Umat^* \uv_0.\label{eqn:LTI_ansatz}
\end{align}
The diagonal elements of the matrix $\Dmat$ contain the complex eigenvalues $d_k$ of the system. Each of these eigenvalues contain the amplification rate $\sigma$ and the frequency $\omega$ of the related mode $d_k  = \sigma_k + \ii \omega_k$ ($\ii=\sqrt{-1}$). The diagonal matrix can be decomposed in amplification rate $\Sigmamat$ and frequency $\Omegamat$, thus $\Dmat = \Sigmamat + \ii \Omegamat$. 
Note that for this linear approach, the non-linearity and non-normality of the Navier-Stokes equations contribute to parameter variations of the linear system. 
The flow is assumed to behave like a linear normal system within the time scale of the filter and all non-linear and non-normal dynamics are represented by variations in  $\sigma$ and  $\omega$. This kind of approach is also pursued in the generalized mean field model of \citet{Luchtenburg.2009}, where the mode interaction via the mean flow is represented by an interaction of linear oscillators trough nonlinear coupling of model parameters.

With the formulation in \eqref{eqn:LTI_ansatz}, the inner product, which forms the elements of the correlation matrix, is simplified to
\begin{align}
\left<u(\xx,t_1),u(\xx,t_2)\right> &= \left<\Umat e^{\Dmat t_1} \Umat^* \uv_0,\Umat e^{\Dmat t_2} \Umat^* \uv_0\right>\\
&= \left<\Umat^* \uv_0,\left(\Umat e^{\Dmat t_1}\right)^* \Umat e^{\Dmat t_2} \Umat^* u_0 \right>\\
&= \left<\Umat^* \uv_0,e^{\Dmat^* t_1} e^{\Dmat t_2} \Umat^* \uv_0 \right>\\
&= \left<\Umat^* \uv_0,e^{(\Sigmamat- i\Omegamat)t_1 + (\Sigmamat + i\Omegamat)t_2} \Umat^* \uv_0 \right>\\
&= \left<\Umat^* \uv_0,e^{\Sigmamat(t_1+t_2) + \ii \Omegamat(t_2-t_1)} \Umat^* \uv_0 \right>.\label{eqn:LTI_corrmat}
\end{align}
An inspection of the exponent in \eqref{eqn:LTI_corrmat} reveals that the inner product only depends on the sum and difference of time steps. 
According to this equation, the velocity fields are projected onto the subspace spanned by the linear operator $\Umat^* \uv_0$ and changes of the inner product are only governed by the eigenvalues of the system. 

Within the context of the correlation matrix, we use the abbreviated nomenclature for the snapshots $\uv_i = \uv(i\deltat)$ and for the projected velocity $\tilde{\uv} = \Umat^* \uv_0$.
The correlation matrix constructed around the neighborhood of snapshots $\uv_0$, yields 
\begin{align}
\Smat_{\mathrm{sub}} &= 
\left(  
\begin{array}[c]{ccc}
\left<\uv_{-1},\uv_{-1}\right> & \left<\uv_{0},\uv_{-1}\right> & \left<\uv_{1},\uv_{-1}\right>\\
\left<\uv_{-1},\uv_{ 0}\right> & \left<\uv_{0},\uv_{ 0}\right> & \left<\uv_{1},\uv_{ 0}\right>\\
\left<\uv_{-1},\uv_{ 1}\right> & \left<\uv_{0},\uv_{ 1}\right> & \left<\uv_{1},\uv_{ 1}\right>\\
\end{array}
 \right) \\
& = 
\left(  
\begin{array}[c]{ccc}
\left<\tilde{\uv},e^{-2 \Sigmamat \deltat} \tilde{\uv}\right> & \left<\tilde{\uv},e^{-\Sigmamat \deltat - \ii\Omegamat \deltat} \tilde{\uv}\right> & \left<\tilde{\uv},e^{-2 \ii\Omegamat \deltat} \tilde{\uv}\right>\\
\left<\tilde{\uv},e^{-\Sigmamat \deltat + \ii\Omegamat \deltat} \tilde{\uv}\right> & \left<\tilde{\uv},\tilde{\uv}\right> & \left<\tilde{\uv},e^{\Sigmamat \deltat - \ii\Omegamat \deltat} \tilde{\uv}\right>\\
\left<\tilde{\uv},e^{2 \ii\Omegamat \deltat} \tilde{\uv}\right> & \left<\tilde{\uv},e^{\Sigmamat \deltat + \ii\Omegamat \deltat} \tilde{\uv}\right> & \left<\tilde{\uv},e^{2 \Sigmamat \deltat} \tilde{\uv}\right>\\
\end{array}
 \right) . \label{eqn:LTI_corr_subset}
\end{align}
The actual properties of this complex expression are highlighted by showing just the factors for $\Sigmamat$ and $\ii \Omegamat$ in the exponents, given as
\begin{align}
\Sigmamat :  
\left[  
\begin{array}[c]{ccc}
-2 & -1 &  0\\
-1 &  0 &  1\\
 0 &  1 &  2
\end{array}
 \right] \deltat 
\ ; \quad
\ii \Omegamat : 
\left[  
\begin{array}[c]{ccc}
 0 & -1 & -2\\
 1 &  0 & -1\\
 2 &  1 &  0
\end{array}
 \right] \deltat. \label{eqn:LTI_corr_subset_exp}
\end{align}
It is perfectly visible that the changes of the correlation matrix along the diagonals are caused by amplification of modes, while changes along the anti-diagonals are caused by the modes' frequency. 
Therefore, the changes of the correlation matrix along the diagonal are related to the modes amplitude and the crosswise changes are caused by the modes phase. 
This was already observed in section \ref{corr_matrix}, where the correlation matrix was compared to the temporal auto-correlation. Note that the diagonals in the figures start at the lower left corner whereas in the formulas they start in the upper left.

As mentioned in section \ref{spectral_POD}, the SPOD and the Fourier transform become equal when the filter operation is extended over the entire sequence and periodic boundary conditions are applied. 
The correlation matrix becomes a Toeplitz (circulant) matrix for this case, and also the dynamics of the entire sequence are described by a linear system. 
The latter point is evidenced by equation \eqref{eqn:LTI_corrmat}. 
For the entire time series to be governed by a linear system, the complete correlation matrix must have the form of the subset shown in equation \eqref{eqn:LTI_corr_subset}. 
Since there can't be any amplification in case of periodic boundary conditions, the correlation matrix is solely defined by the frequencies of the modes ($\Omegamat$). 
Thus, the matrix is constant along the diagonals, which is the presumed shape.

\begin{figure}
	\centering
			\includegraphics{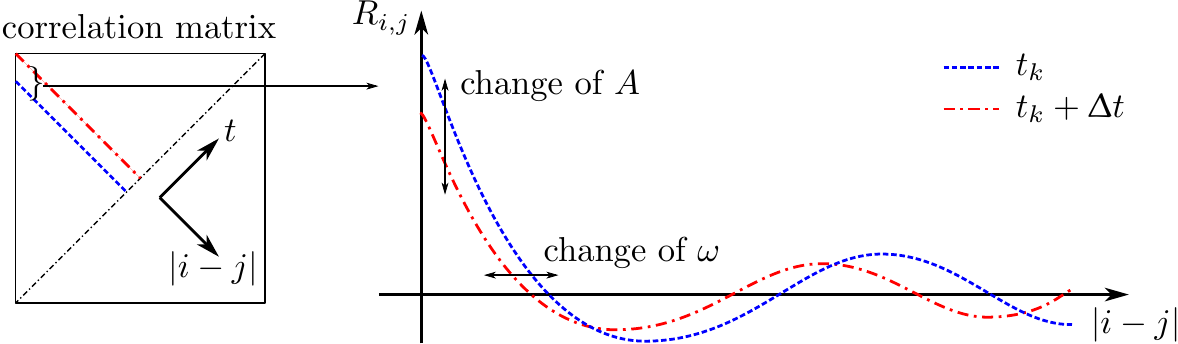}
	\caption{Schematic that illustrates the influence of parameter changes on the correlation matrix anti-diagonals. The correlation matrix on the left indicates the orientation of the diagonals (different times $t$) and the anti-diagonals (different time delays $\tau = |i-j|\deltat$). The parameters $A$ and $\omega$ are specified in \eqref{eqn:anal_signal}.}
	\label{fig:corr_change_sketch}
\end{figure}

Recall that the SPOD filter operation acts along the diagonals of the correlation matrix. 
The excursion to system dynamics indicates that the SPOD puts a constraint on the temporal variation of the mode coefficients. 
To clarify these properties we describe the coefficient and its derivative by an analytical signal
\begin{align}
\widetilde{b}(t) = A(t) e^{\ii\phi(t)} \ ;\quad \frac{\partial \widetilde{b}}{\partial t} = \left[\sigma(t) + \ii \omega(t)\right]\widetilde{b}(t)  \label{eqn:anal_signal}.
\end{align}
In this framework, changes along the diagonal of the subset \eqref{eqn:LTI_corr_subset_exp} are caused by temporal variations of $A$ or $\omega$. 
This behavior is sketched in figure \ref{fig:corr_change_sketch}, where the corresponding changes of the correlation matrix anti-diagonal are depicted.
There, changes of the amplitude cause an overall scaling and changes of the frequency move the zero crossings.
The SPOD filter \eqref{eqn:SPOD_filter} smooths along the diagonals and therefore it equalizes the anti-diagonals of consecutive time steps. 
Thus, the two curves in figure \ref{fig:corr_change_sketch} move closer together, which limits the rate of change of the amplitude $\partial A/ \partial t$ and the frequency $\partial\omega/ \partial t$. 

The exact spectral properties of the mode coefficients might not be directly concluded from the considerations in this section, but they become obvious in the SPOD results. 
They can be summarized as follows:
The low-pass filter applied in equation \eqref{eqn:SPOD_filter} defines a certain spectral bandwidth. 
The filter response and cut-off frequency can be obtained from the coefficients $g_i$. 
The individual SPOD coefficients $b_i$ feature low-pass filtered amplification rates $\sigma$ and band-pass filtered frequencies $\omega$. 
For higher SPOD modes the effect of the filter becomes less pronounced. 
This is plausible, since the filter is applied to the entire correlation matrix and not as a constraint for the single modes.

\subsection{Identification of coupled modes}\label{SPOD_mode_link}

\begin{figure}
	\centering
			\includegraphics{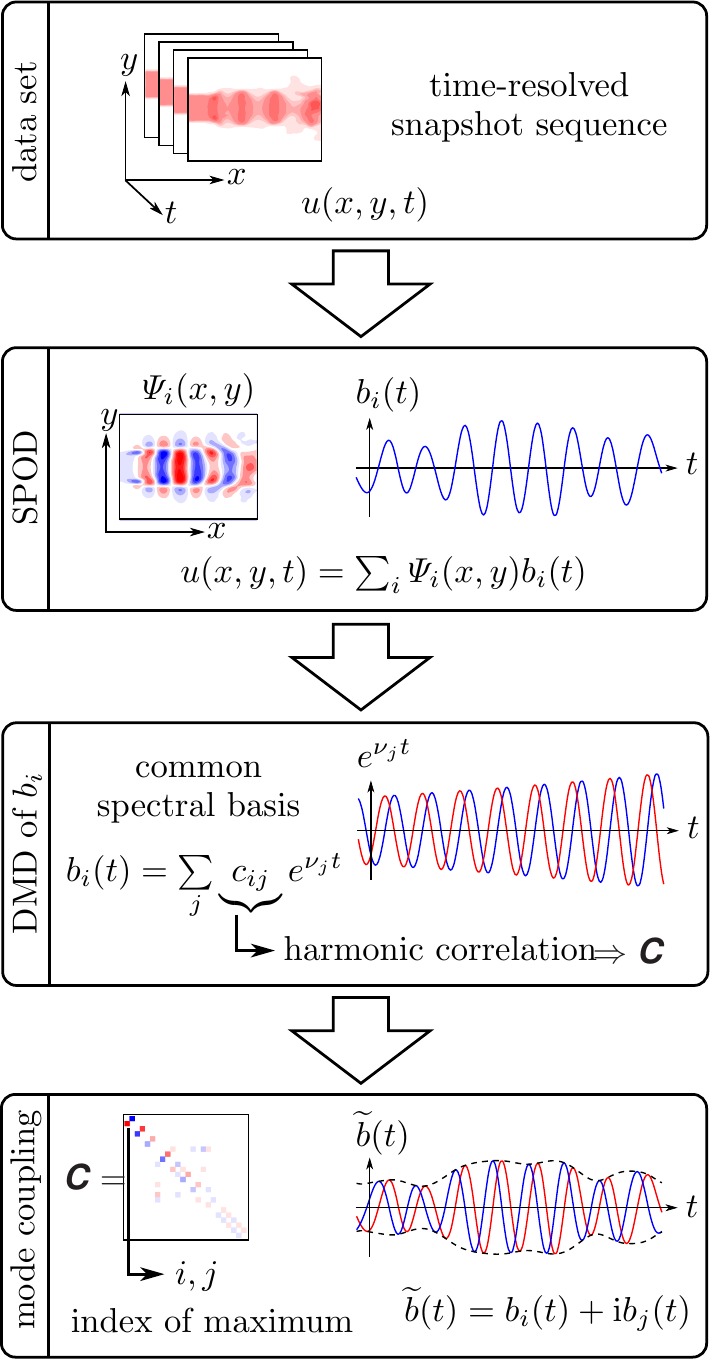}
	\caption{Schematic illustrating the main steps towards the identification of coupled modes (red and blue lines indicate real and imaginary parts of an analytic signal). The data displayed here were derived from measurements of a forced turbulent jet (see also figure \ref{fig:corr_mat}).}
	\label{fig:mode_coupling_sketch}
\end{figure}

One crucial point in POD and SPOD is the identification of linked modes. 
Assuming the presence of periodic coherent structures, their dynamics are described by a pair of modes, analogue to the sine and cosine in the Fourier space or the real and imaginary part of DMD modes. 
They constitute the real and imaginary part of the analytical coefficient in equation \eqref{eqn:anal_signal}. 
The coupling of such a mode pair is not given by the SPOD and it has to be identified a posteriori. 
Coupled modes typically show a similar amount of energy and pair in the POD spectrum \citep{Oberleithner.2011,Oberleithner.2014}. 
For more complex dynamics with multiple energetic modes, the pairs are not easily identified and visual inspection of Lissayous figures and spatial modes is required.  
This manual procedure is cumbersome and by no means objective. 

To provide an alternative, we propose an unbiased approach that gives a quantitative measure of the dynamic coupling of individual modes.  
Based on a DMD of the temporal coefficients, we derive a relation of the spectral proximity of the SPOD modes. 
The general procedure is schematically outlined in figure \ref{fig:mode_coupling_sketch}. 
From the depicted operations, the DMD and the mode coupling are discussed in this section.
The Fourier decomposition of the coefficients may similarly provide a basis to compare the coefficients, but the DMD turned out to be more reliable for this task.

For the DMD of the SPOD coefficients, their temporal evolution is assumed to be governed by a linear operator $\Tmat$
\begin{align}
\bv(t+\deltat) = \Tmat \bv(t).
\end{align}
To approximate this operator, the SPOD coefficients are arranged in two matrices $\XX = [\bv(0) \ \bv(\deltat) \ ... \ \bv((N-2)\deltat)]$ and $\YY = [\bv(\deltat) \ \bv(2\deltat) \ ... \ \bv((N-1)\deltat)]$ (following the notation of \citet{Tu.2014}). Hereafter, the operator is given by
\begin{align}
\Tmat = \YY \XX ^{-1},
\end{align}
where $\XX ^{-1}$ is the Moore-Penrose pseudoinverse of $\XX $. 
Alternatively, this can be solved as a least squares problem, minimizing $\left\|\Tmat\XX-\YY\right\|$. 
To reject measurement noise in the identification procedure, only the ``physical part'' of the SPOD modes is considered for the calculation of the operator $\Tmat$. 
That means, only the modes with acceptable signal to noise ratio should be considered.
Therefore, the number of retained modes is calculated from the energy resolved by the SPOD, truncated after ${N_c}$ modes, with
\begin{align}
\mathcal{E}(N_c) = \frac{\sum_{k=1}^{N_c}{\mu_k}}{\sum_{k=1}^{N}{\mu_k}}.
\end{align}
In the examples shown later the modes are truncated around $\mathcal{E}(N_c) = 0.95$. 
This value depends on the signal to noise ratio of the considered measurement, which can be estimated from the POD spectrum \citep{Raiola.2015}. 
Note that the number of retained modes increases for wider SPOD filters and corresponding flatter SPOD spectra ($\mu_j$).

The DMD modes are obtained by the eigen-decomposition of matrix $\Tmat$ as 
\begin{align}
\Tmat \cv_i = \nu_i \cv_i.
\end{align}
The eigenvalues $\nu_i$ comprise the frequencies $\omega_i$ and amplification-rates $\sigma_i$ of the operator $\Tmat$ and are given by the logarithm of the eigenvalue $\ln(\nu_i)/\deltat = \sigma_i + \ii \omega_i$. 
The eigenvectors $\cv_i$ hold the relative spectral content of the single SPOD coefficients with respect to these frequencies. 
More precisely, the element $c_{i,j}$ of vector $\cv_i$ is the spectral content of the single mode coefficient $b_j$ with respect to $\nu_i$. 
The actual modal representation is given by
\begin{align}
b_j(t) = \sum_{i=1}^{N_c}{c_{i,j} e^{(\sigma_i + \ii \omega_i)t}}.
\end{align}
It must be noted that this decomposition is only exact if $N_c = N$, whereas in the current approach $N_c<N$. 
Nevertheless, the decomposition gives a common spectral basis, which allows the ranking of spectral similarity of the temporal coefficients $\bv(t)$.
The developed proximity measure is given by 
\begin{align}
C_{i,j} = \Imag \left\{\sum_{k=1}^{N_c}{c_{k,i} c_{k,j}^*\sgn\left(\Imag \left(\nu_k\right)\right)}\right\},
\label{eqn:mode_link}
\end{align}
where the coefficients are normalized to $(\cv_i,\cv_i)=1$. The sign ($\sgn$) in this expression accounts for the conjugate pairs that appear in the DMD spectrum (mirrored at the real axis). 

For two modes to be coupled, they must have a similar spectral content, which is either shifted a quarter period forward or backward. 
This implies a positive or negative imaginary part of the harmonic correlation \eqref{eqn:mode_link}, respectively, and coupled modes appear as peaks in the matrix $\Cmat$.
Hence, the row and column indices of the maximum of $\Cmat$ identify the first coupled SPOD modes.
The corresponding row and column in $\Cmat$ are then set to zero and the next lower maximum is identified.
This procedure is repeated  until all modes are paired.
It has to be noted that this approach also creates weakly correlated and possibly unphysical mode pairs.

Together with the identification of coupled modes, the procedure gives an average frequency of the coherent structure represented by the mode pair.
Therefore, the eigenvalues $\nu_k$ of the matrix $\Tmat$ are sorted in descending order with respect to their content for the identified mode pair $\widetilde{c}_k = c_{k,i}^2 + c_{k,j}^2$.
The frequency is given as the weighted sum of the eigenvalues
\begin{align}
f =  \frac{\sum_{i=1}^n{\Imag \left\{\ln(\nu_i)\right\} \tilde{c}_i }}{2 \pi \deltat \sum_{i=1}^n{\tilde{c}_i}}.
\label{eqn:mode_link_freq}
\end{align}
The weighting accounts for the relative energy content of a mode pair with respect to the single frequencies. In fact, only the most relevant eigenvalue ($n=1$) can be picked to determine the frequency, but for practical application it is recommended to use more than one eigenvalue as noise may corrupt them. For the examples discussed in the next chapter, we used an average over three eigenvalues ($n=3$) to get  accurate results. 

The coupled SPOD modes are considered as {\itshape{one}} complex mode (see equation \eqref{eqn:anal_signal} and figure \ref{fig:mode_coupling_sketch}) similar to the Fourier mode. The relative energy content of the identified modes is computed as 
\begin{align}
K =  \frac{\mu_i + \mu_j}{\sum_{k=1}^{N_c}{\mu_k}},
\label{eqn:mode_link_energy}
\end{align}
where $i$ and $j$ again refer to the indices of the coupled SPOD modes.

\section{Applications to experimental data}\label{sec:application}

In this chapter the SPOD is applied to three different data sets. 
All three examples originate from very different engineering problems, demonstrating the capability and broad applicability of SPOD. 
We consider the flow of a swirl-stabilized combustor, the wake of an airfoil with Gurney flap, and  the flow field of a sweeping jet generated with a fluidic oscillator. 
All three flows were recorded with the same PIV measurement system. 
It consists of a Photron Fastcam SA 1.1 high-speed camera (1Mpixel at 2.7kHz double frame) and a Quantronix Darwin Duo laser (30mJ at 1kHz). 
The PIV data were processed with PIVview (PIV{\textit{TEC}} GmbH) using standard digital PIV processing \citep{Willert.1991} enhanced by iterative multigrid interrogation with image deformation \citep{Scarano.2002},\citep[pp. 146-158]{Raffel.2007}. 
Analyzing the present data sets with existing POD, DFT or DMD approaches caused some difficulties. 
As will be demonstrated, the SPOD is able to handle these shortcomings. 
The DMD and the DFT equally suffer from the restriction of the modes to narrow frequency bands, therefore we limit the following presentation to DFT. 
This choice is particularly handy as the DFT is a limiting case of the SPOD.

\subsection{Swirling jet undergoing vortex breakdown}\label{application_swirl}

\begin{figure}
	\centering
	\includegraphics{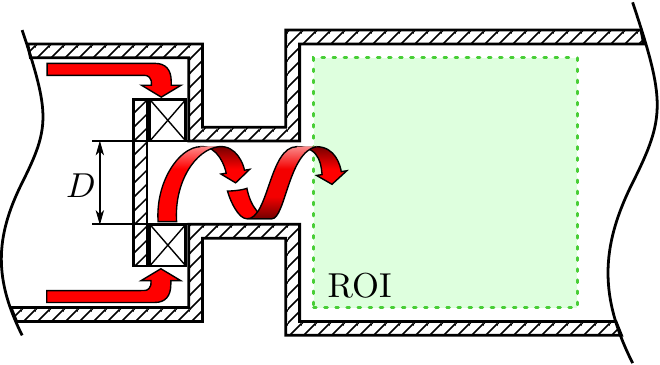}
	\caption{Experimental setup of the swirl stabilized combustor. Air enters from the left, passes a swirl generator and exits into the combustion chamber. Flow field measurements with PIV are conducted in the meridional plane as indicated by dashed square (ROI).}
	\label{fig:swirl_setup}
\vspace{11pt}
	\centering
			\includegraphics{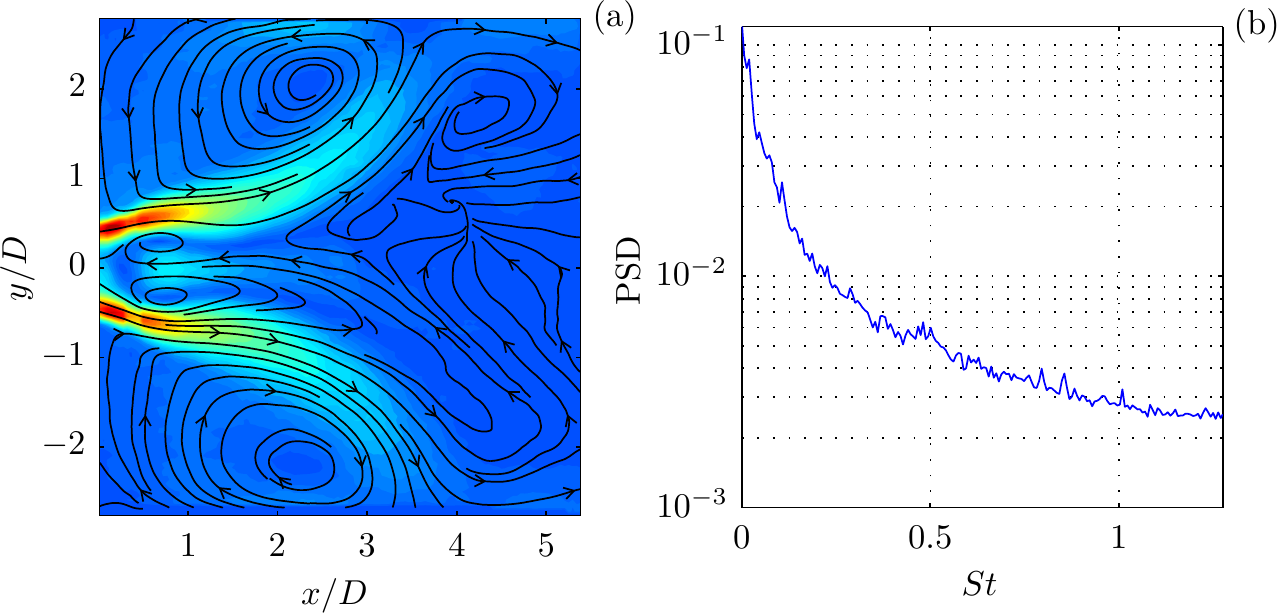}
	\caption{Swirl-stabilized combustor flow: Time-averaged flow field depicted by (a) contours of velocity magnitude and streamlines, and (b) spatially-averaged power spectral density.}
	\label{fig:swirl_mean_flow}
\end{figure}

At first, we consider the flow field of a swirl-stabilized combustor. 
Swirling jets are widely used in the gas turbine industry due to their capability of obstacle-free flame anchoring and enhanced mixing. 
The experimental setup to study these flows is sketched in figure \ref{fig:swirl_setup}.
Swirl is generated by injecting fluid tangentially into a mixing tube that terminates in the combustion chamber. 
The cylindrical-shaped chamber is made of quartz glass to allow optical access for PIV. Flow measurements are conducted in the meridional section as indicated in the schematic. 
The case investigated here is non-reacting at a Reynolds number of 58 000 based on the nozzle diameter $D$ and the bulk velocity at the nozzle exit. 
Additional details about the experimental setup can be found in \citet{Reichel.2015}.

\begin{figure}
	\begin{flushleft}
	(a)
	\end{flushleft}
	\vspace{3pt}
	\centering
			\includegraphics{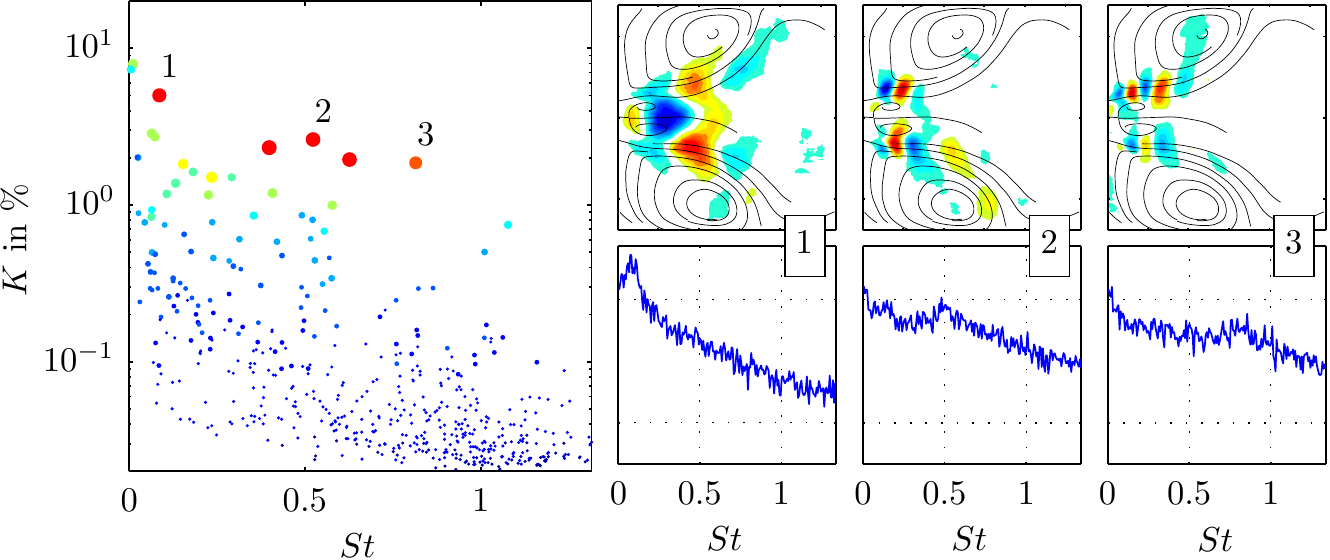}
	\begin{flushleft}
	(b)
	\end{flushleft}
	\vspace{3pt}
	\centering
			\includegraphics{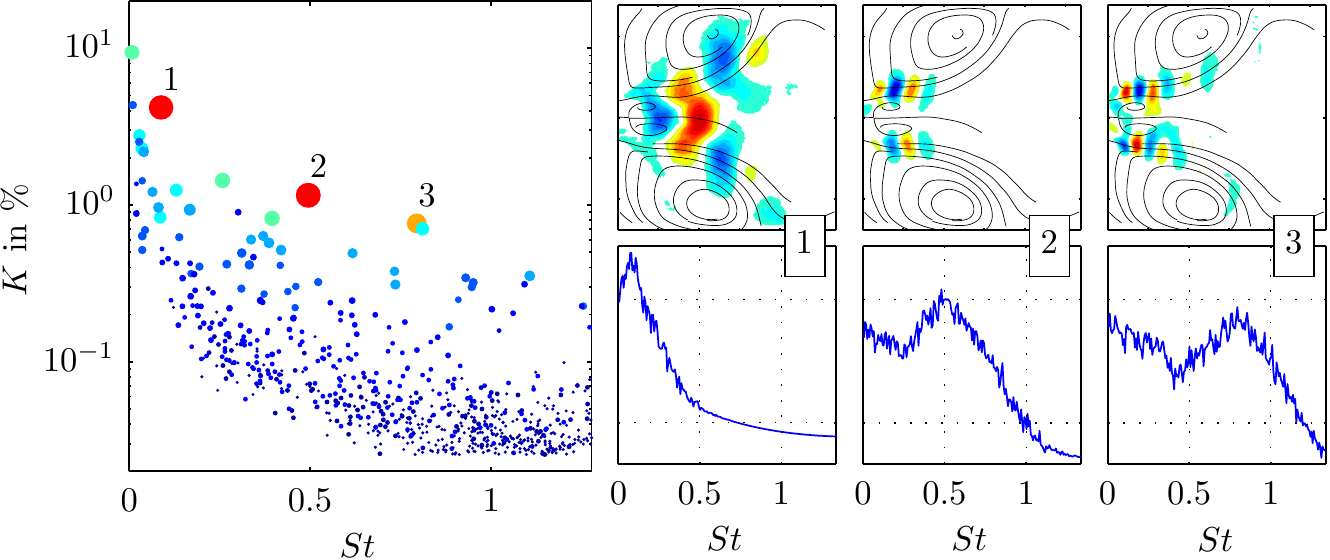}
	\begin{flushleft}
	(c)
	\end{flushleft}
	\vspace{3pt}
	\centering
			\includegraphics{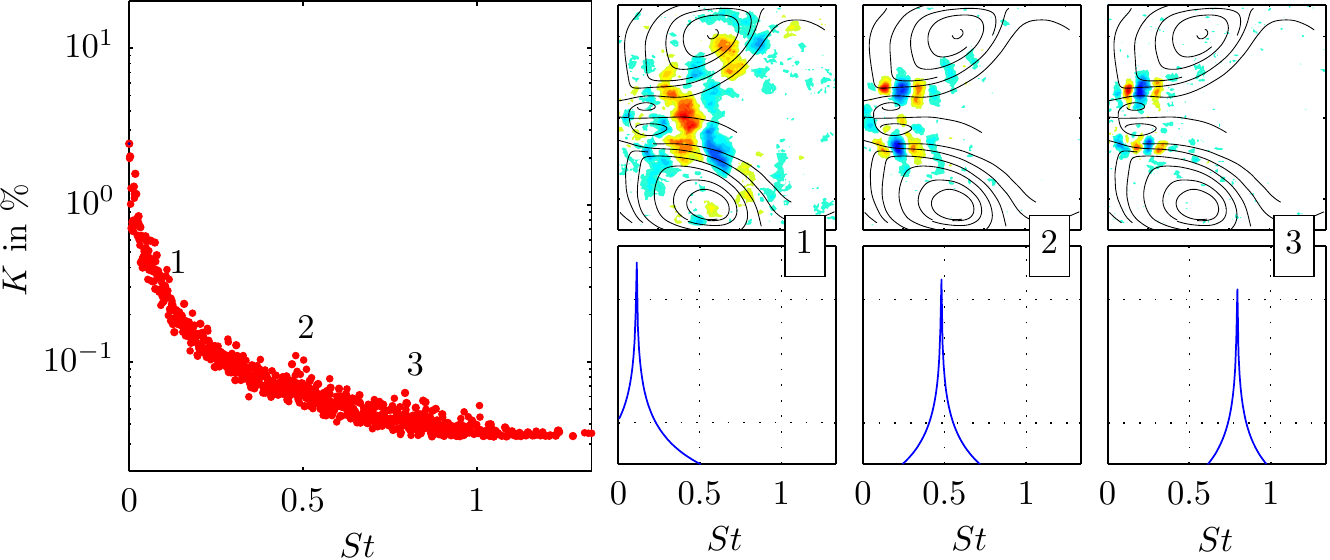}
	\caption{ Swirling jet: Results from SPOD for different filter lengths (a) $N_f= 0$ (POD), (b) $ N_f = 10$ (SPOD), and (c) $N_f=2000$ (DFT). 
	For every filter length the SPOD spectrum is displayed as scatter plot (left), where a single dot indicates one mode pair (size and color $C_{i,j}$ in \eqref{eqn:mode_link}).
	For three selected pairs the spatial modes (upper row) and PSD of the temporal coefficient (lower row) are depicted. 
	They are indicated by numbers in the SPOD spectrum, as well as between the small mode plots.}
	\label{fig:swirl_spectrum}
\end{figure}

The mean flow field and the spatially-averaged power spectral density (PSD) is depicted in figure \ref{fig:swirl_mean_flow}, with the Strouhal number based on the same length and velocity scale as the Reynolds number ($\Stn = f D / u_{\mathrm{bulk}}$). 
The flow exhibits a strong recirculation zone in the center, surrounded by an annular, strongly diverging jet. 
The cross-sectional jump at the combustion entrance creates an additional external recirculation zone between the jet and the confining walls. 
The spectral content of the flow is spread over all time scales and it decreases with increasing Strouhal number. 
The spectrum gives no indication of any dominant coherent structure even though these flows typically feature helical global modes \citep{Oberleithner.2011}.

Figure \ref{fig:swirl_spectrum} (a), (b), and (c) illustrate the results from the SPOD for the filter lengths $N_f=0$, 10, and 2000, respectively. 
Note that the limiting cases $N_f=0$ and $N_f=2000$ produce results equivalent to those obtained with classical POD and DFT respectively, while the case in between represents the SPOD. 
Hence, this particular presentation concisely demonstrates the difference between POD, SPOD and DFT. 

Each of the three cases in figure \ref{fig:swirl_spectrum} show the so-called SPOD spectrum, where every mode pair is represented by a single dot. 
The size and color of the dots indicate the harmonic correlation of a mode pair $C_{i,j}$, according to equation \eqref{eqn:mode_link}.
The frequency of a mode pair is determined according to \eqref{eqn:mode_link_freq} and the energy from the two eigenvalues relative to the total energy from \eqref{eqn:mode_link_energy}.
On the right side of every case in figure \ref{fig:swirl_spectrum}, three spatial modes $\Psi_i(\xx)$ and the corresponding temporal coefficients $b_i(t)$ are posed above each other. 
The spatial modes are visualized by the crosswise velocity component (in $y$-direction) together with streamlines of the time-averaged flow. 
They are numbered likewise in the SPOD spectrum and between the small mode plots.
The plots are given without axis labels to allow a compact representation of the data, the section is the same as for the mean flow shown in figure \ref{fig:swirl_mean_flow}(b). 
The time coefficients are represented by their power spectral density, where the time series is split into five (50\% overlapping) sections, which are Fourier transformed and averaged. 
The horizontal dotted lines in the PSD plots indicate a spacing of three orders of magnitude (1000) and the $y$-axis is scaled logarithmically.
The spectral averaging was also applied for the power spectra shown in figure \ref{fig:swirl_mean_flow}(b) and in the subsequent PSD plots.

The POD (figure \ref{fig:swirl_spectrum}(a)) yields a broad spectrum of modes, where modes at lower Strouhal numbers have more energy. 
There are several modes with high harmonic correlation (diameter and color of the points),  and high energy contents $K$. 
The spatial shape of the low-frequency mode (label 1) shows clear spatial symmetry and a limited spectral bandwidth ($\Stn \approx 0.1$).   
This mode is frequently observed in swirl-stabilized combustors and it is associated with a global hydrodynamic instability \citep{Terhaar.2015b}. 
From the four additional outstanding modes between $\Stn=0.3$ and $\Stn=0.8$ we pick two for further investigation. 
Their spatial structures show no clear symmetries and indicate mixtures of several spatial wavelengths. 
Accordingly, the mode spectra are broad and show only a slight hump at the frequencies indicated by the SPOD spectrum. 
The other modes around $\Stn=0.5$, which are not shown here, show similar spatial and spectral content. 

Overall, the POD indicates the presence of a single mode at low frequency, together with other coherent structures that are not properly resolved. 
The first most energetic (not inspected here) corresponds to a low frequency, large wavelength fluctuation, as indicated by the SPOD. 
Such slow changes of the mean flow are usually named shift modes \citep{Luchtenburg.2009,Hosseini.2015}.
In this particular case the shift mode stems from weak movements of the vortex breakdown bubble.

Consider now the SPOD in figure \ref{fig:swirl_spectrum}(b), with a filter length $N_f = 10$; 
From the SPOD spectra we identify three peaks at $\Stn=0.09$, 0.5 and 0.8.
The first mode is the same as the one already identified by the POD, but its spectral content at higher frequencies is reduced. 
It describes a single-helical structure in the wake of the recirculation zone.
The second identified mode exhibits a broad spectral peak at $St=0.5$. 
Its spatial structure and Strouhal number match the global mode identified by \citet{Oberleithner.2011}. 
It is a single-helical mode linked to the precessing motion of the recirculation zone. 
The spatial structure of mode three suggests a double-helical shape. 
It is not a harmonic of the second mode, as their frequencies are not related. 

When the filter size is extended to its maximum, we get the the DFT (figure \ref{fig:swirl_spectrum}(c)) and the SPOD spectrum converges to the averaged PSD (figure \ref{fig:swirl_mean_flow}(b)). 
The temporal coefficients converge to sines and cosines and all mode pairs show full harmonic correlation (uniform dot size in the SPOD spectrum). 
Since the selection based on harmonic correlation is impossible, we resort to the frequencies already known from the SPOD at $N_f=10$. 
The spatial structures resemble the ones from figure \ref{fig:swirl_spectrum}(b), but they are corrupted by noise. 
Moreover, the spatial symmetries are no longer as clear as for the $N_f=10$ case.
Note that the corresponding  spectral peaks are broadened due to the averaging procedure, which is applied here only for consistency. 

From this first example, we can point out some striking features of the SPOD. 
The SPOD is able to separate coherent fluctuation from stochastic turbulent fluctuations even though they both have the same energy contents (see the SPOD spectrum in figure \ref{fig:swirl_spectrum}(c)). 
The classical POD yields partially mixed structures that cannot be assigned to distinct flow phenomena, whereas the SPOD properly separates these structures and identifies them from harmonic correlations.
The DFT instead shows the same structures at the identified frequencies, but they are corrupted with noise and the method itself would give no clue about the frequencies of interest. 

The structures identified with the SPOD may also be found with the POD if the decomposition is applied to a subsection of the measured domain. 
Moreover, the exploitation of spatial symmetries prior to the POD decomposition usually provides good results for this type of flows \citep{Terhaar.2015c}. 
Nevertheless, these alternative approaches would require prior knowledge of the shape or spatial extent of the structures, whereas the SPOD requires none of these. 

All modes identified by the SPOD show clear spatial symmetries and distinct spectral peaks.
The frequency and shape of the first mode coincide well with previous experimental observations in swirl-stabilized combustors \citep{Terhaar.2015b}. 
The second mode is very similar to the one observed in unconfined swirling jets \citep{Oberleithner.2011}.
However, the presence of these different modes in a single flow configuration raises the question about their physical relevance.
To deal with this issue, we conducted a linear stability analysis of the underlying mean flow, following the procedure outlined by \citet{Oberleithner2015a}. 
This analysis similarly delivered three unstable modes whose frequencies and azimuthal and axial wavenumbers match the SPOD modes surprisingly well. 
To limit the scope of this paper the analysis is not further detailed here. 
One important parameter of the SPOD is the filter size $N_f$, which is twice the period of the second mode. 
The experiences gained throughout the first application show that a filter size of one to two periods of the mode of interest gives the best results.

\subsection{Airfoil with Gurney flap}\label{application_flap}

\begin{figure}
	\centering
	\includegraphics{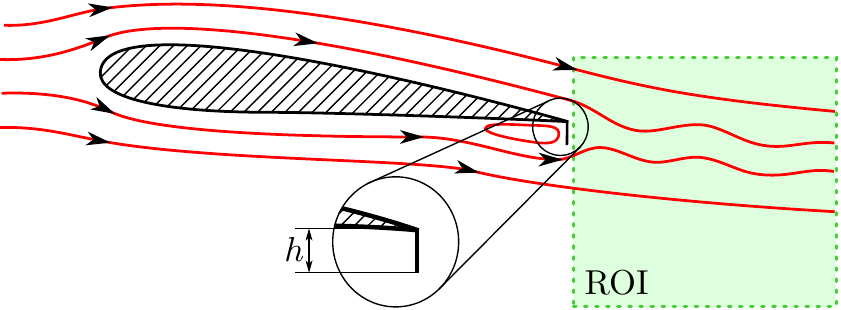}
	\caption{Schematic of the airfoil equipped with a Gurney flap at the trailing edge. Streamlines indicate the surrounding flow and the vortex upstream of the flap. The measured section (ROI) is a streamwise cut in the wake of the airfoil, capturing the periodic shedding behind the flap.}
	\label{fig:flap_setup}
\vspace{11pt}
	\centering
			\includegraphics{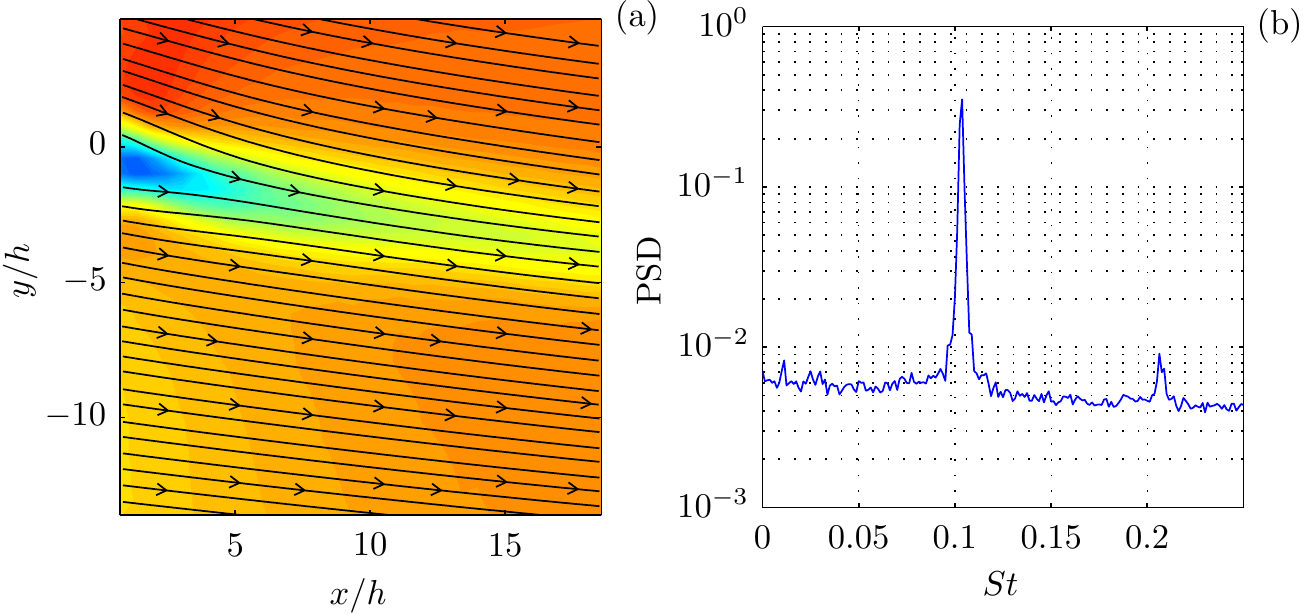}
	\caption{Wake of the airfoil with Gurney flap: Time-averaged flow field depicted by (a) contours of velocity magnitude and streamlines, and (b) spatially-averaged power spectral density.  The origin of the coordinate system is located at the trailing edge.}
	\label{fig:flap_mean_flow}
\end{figure}

The second flow configuration considered here is the flow behind a pitched airfoil equipped with a Gurney flap. 
The experimental setup is shown in figure \ref{fig:flap_setup}. 
It illustrates the working principle of the Gurney flap deployed at the pressure side of the airfoil. 
The flap creates additional lift (and drag), which can be used to locally control varying loads on large wind turbine airfoils \citep{Bach.2015b,Bach.2014}. 
The flow features around the Gurney flap are characterized by a single vortex that develops upstream of the flap and periodic shedding in its wake. 
The vortex upstream of the flap continuously grows up to a critical size, then it is shed into the wake, and it starts growing again. 
Here, a FX 63-137 airfoil  at $5^{\circ}$ angle of attack is investigated at a Reynolds number of 180 000 based on chord length. 
The reference length for the Strouhal number is the flap height, which is 3.6\% of the chord. 
The measured region comprises only the wake of the airfoil capturing the dominant vortex shedding. 
More details about the experimental setup can be found in a preceding publication of these data \citep{Bach.2015}. 
The Strouhal number in the following results is calculated with the flap height $h$ and the free stream velocity.

The mean flow shown in figure \ref{fig:flap_mean_flow}(a) reveals a velocity deficit in the wake of the Gurney flap, which generates the vortex shedding. 
The PSD (figure \ref{fig:flap_mean_flow}(b)) indicates strong oscillations at $St=0.105$ with a weak higher harmonic.
In a previous investigation it was found that the vortex, which is shed from upstream of the flap causes an alteration of the periodic vortex shedding behind the flap \citep{Troolin.2006}. 
Hot-wire measurements in the wake of the Gurney flap supported this assumption. 
The combination of a strong periodic flow pattern and the intermittent short-time events provides a formidable benchmark for the SPOD. 

\begin{figure}
	\begin{flushleft}
	(a)
	\end{flushleft}
	\vspace{3pt}
	\centering
			\includegraphics{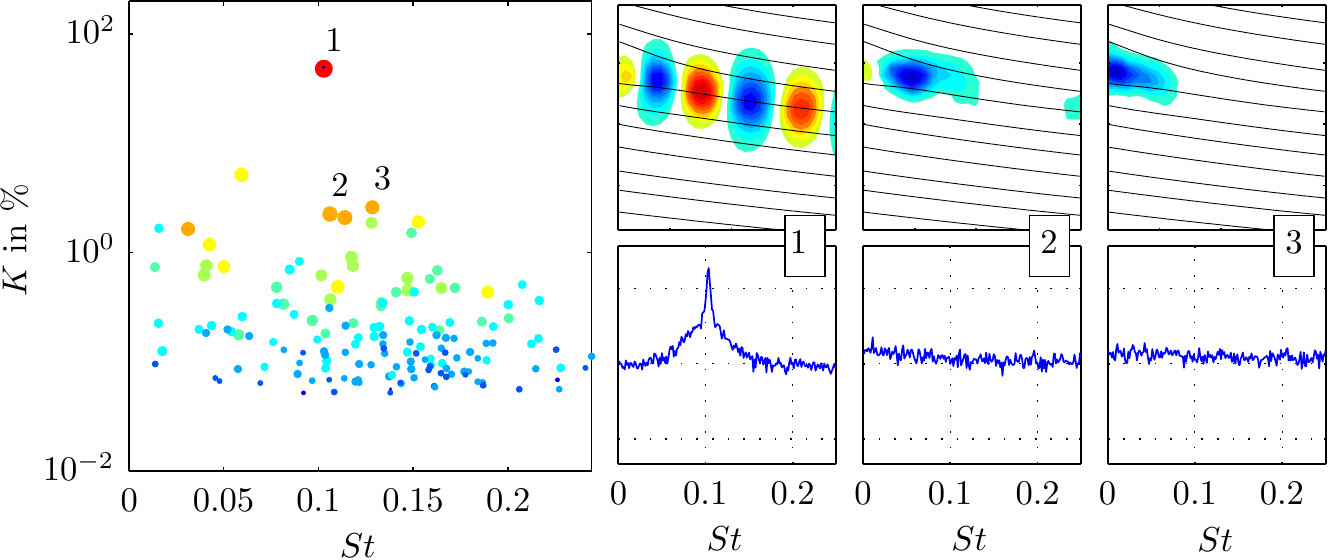}
	\begin{flushleft}
	(b)
	\end{flushleft}
	\vspace{3pt}
	\centering
			\includegraphics{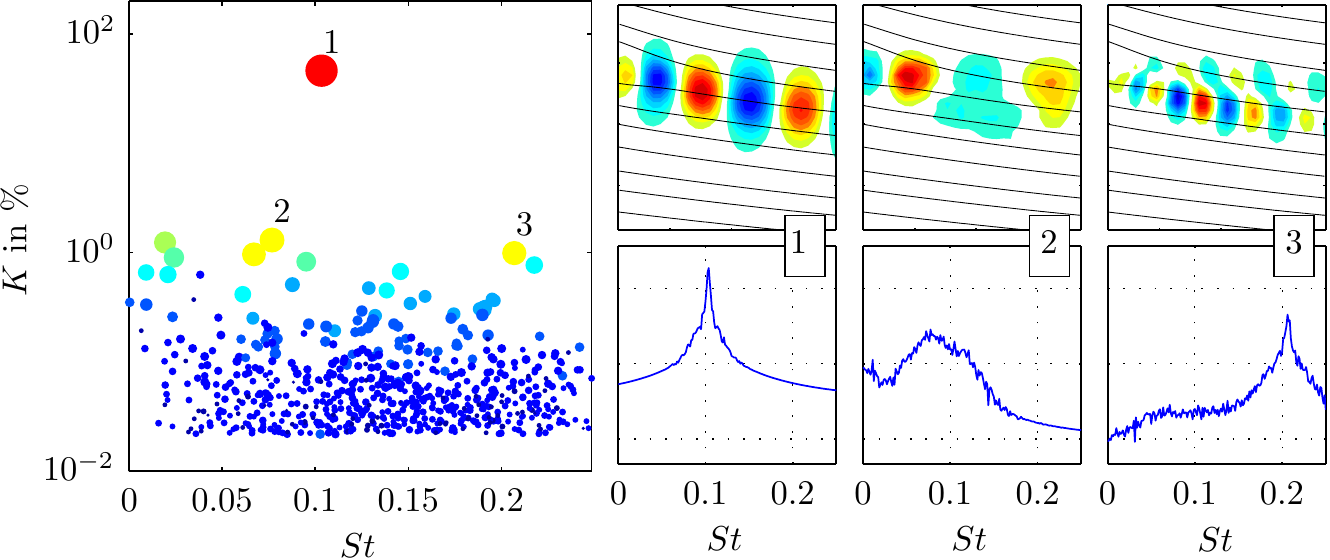}
	\begin{flushleft}
	(c)
	\end{flushleft}
	\vspace{3pt}
	\centering
			\includegraphics{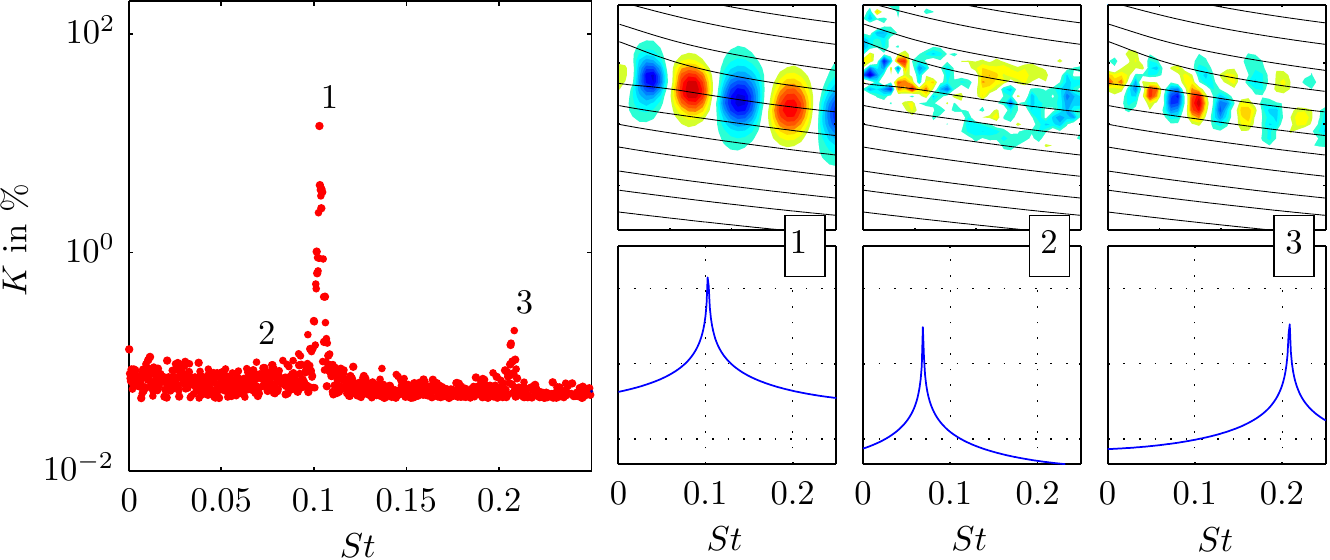}
	\caption{Airfoil with Gurney flap: Results from SPOD for different filter lengths (a) $N_f= 0$ (POD), (b) $ N_f = 15$ (SPOD), and (c) $N_f=2000$ (DFT). For every filter length the SPOD spectrum is displayed as scatter plot (left), where a single dot indicates one mode pair (size and color $C_{i,j}$ in \eqref{eqn:mode_link}). For three selected pairs the spatial modes (upper row) and PSD of the temporal coefficient (lower row) are depicted. They are indicated by numbers in the SPOD spectrum, as well as between the small mode plots.}
	\label{fig:flap_spectrum}
\end{figure}

The presentation of the decomposition with the different methods is organized in the same way as for the previous example. 
The classic POD decomposition is shown in figure \ref{fig:flap_spectrum}(a).
The vortex shedding is represented by the most energetic POD mode with the highest harmonic correlation.
The remaining modes show weak harmonic correlations and no distinct peak in the SPOD spectrum. 
The modes labeled 2 and 3 exhibit a broad spectral content with a spatial extent limited to the vicinity of the flap. 
There are additional modes with similarly compact spatial extent located further downstream. 
These compact modes describe the intermittent alteration of the vortex shedding during the passage of the single vortex that was generated upstream of the Gurney flap. 
An inspection of their time coefficients (not shown) reveals that these modes are only active one after another during one shedding period. 
Depending on the phase lag between the natural oscillation and the shedding of the upstream vortex, the developing wake vortex is either strengthened or weakened. 
The convection of this altered vortex is described by the spatial series of modes.
This behavior is indicated by a complex interaction of these modes and the periodic shedding modes, which is hard to identify in the POD expansion.

The SPOD yields a much clearer set of modes (figure \ref{fig:flap_spectrum}(b)). 
In addition to the shedding mode, the SPOD also uncovers three other modes, which are offset from the rest.
The two modes that appear at similar frequencies capture the alteration of the vortex shedding during the passage of the single vortex.
Their mode shape is similar to the shedding mode, but with larger spatial wavelengths and lower frequencies  (see mode 2 in figure \ref{fig:flap_spectrum}(b)). 
The interaction of the upstream vortex with the vortex shedding increases the vortex size and thus the wavelength in the wake. 
Assuming a constant convection speed in the flow, this mode consequently has a lower frequency. 
In case of the SPOD the alteration of the vortex shedding is captured by a single mode (pair) and is thus much easier to interpret. 
The third mode represents the second harmonic of the vortex shedding with a clear spectral peak and clean spatial mode with twice the wavelength of the shedding mode. 
This higher harmonic is completely missed in the POD. 
The SPOD filter size $N_f$ is equivalent to three shedding periods, which is approximately equal to the traveling time through the measurement domain.

The DFT shown in figure \ref{fig:flap_spectrum}(c) reproduces the spectrum shown in figure \ref{fig:flap_mean_flow}(b). 
The natural mode and its higher harmonic can be identified from the spectral peaks. 
The corresponding mode shapes are similar to the SPOD, although the higher harmonic is corrupted with noise, resulting in a fragmented spatial mode. 
The DFT at the frequency of the second SPOD mode gives no indication of the structure identified before and the vortex-vortex interaction is completely missed. 
This is attributed to the fact that this phenomenon is highly intermittent with varying frequencies and amplitudes, which cannot be represented by a single-frequency mode. The same dilemma applies for the DMD. 

For this example, the SPOD has shown its ability to separate dynamics with similar spatial structures and frequencies but very different energies. 
The spectral proximity and spatial similarity of the involved dynamics impede the application of POD. 
The modulation of the natural vortex shedding is represented by a natural mode with several intermittent modes. 
The DFT, however, with its single frequency modes does not capture the modulation of the shedding at all.  
The frequency constraint imposed by the SPOD is sharp enough to split the natural shedding from the modulation and soft enough to allow for frequency and amplitude variations. 
Hence the SPOD again gives easy access to dynamic features of the flow, which cannot be found with other common methods of similar algorithmic complexity. 
There may be feature tracking approaches capable of identifying the dynamics in this case, but they usually require more computational effort and might not be as versatile as the SPOD.

\subsection{Fluidic oscillator}\label{application_fluidix}

\begin{figure}
	\centering
	\includegraphics{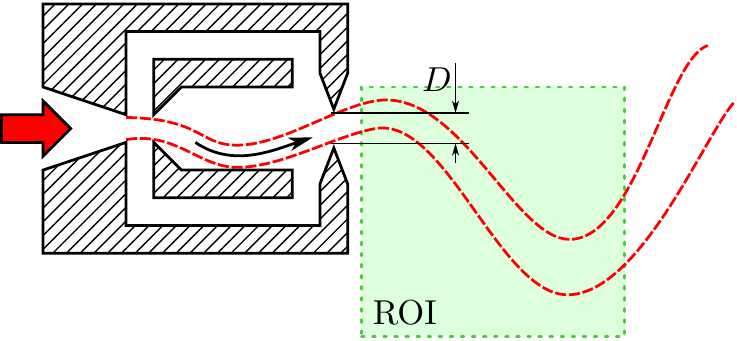}
	\caption{Schematic of the experimental setup with the fluidic oscillator. Air enters from the left, passes the oscillator and exits into the unconfined ambient air. The angle of the jet leaving the oscillator sweeps periodically up and down. The measured region (ROI) captures the meridional plane of the jet's near field. The oscillator has a square nozzle, hence the thickness of the jet normal to the plane is also $D$.}
	\label{fig:fluidix_setup}
\vspace{11pt}
	\centering
			\includegraphics{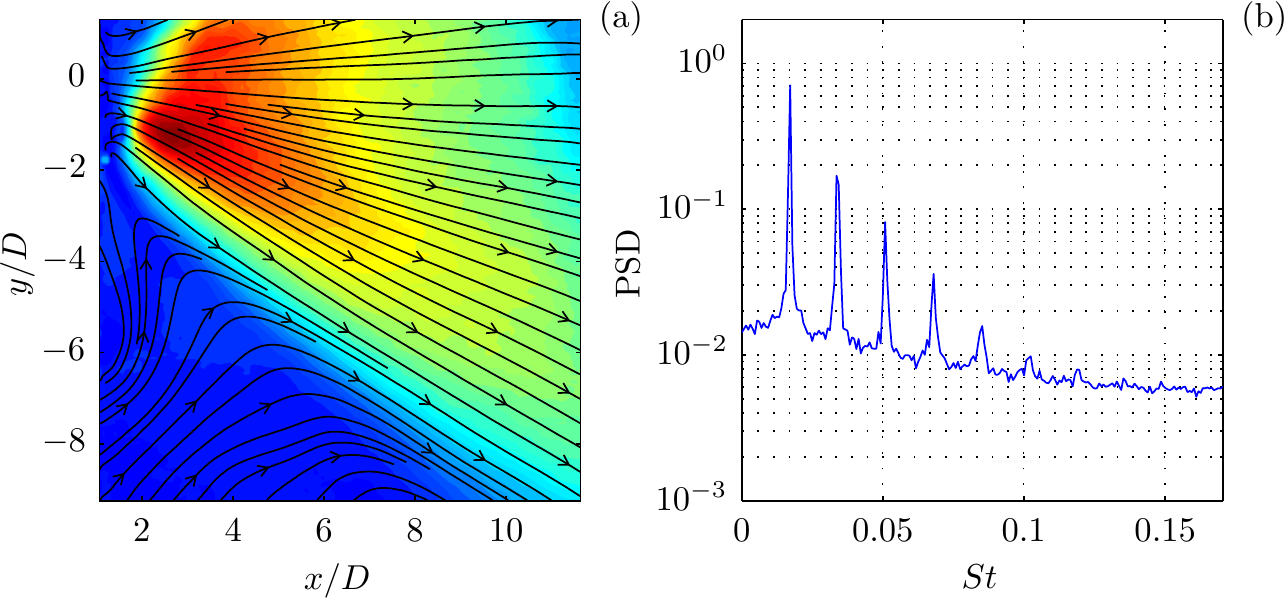}
	\caption{Fluidic oscillator: Time-averaged flow field depicted by (a) contours of velocity magnitude and streamlines, and (b) spatially-averaged power spectral density.}
	\label{fig:fluidix_mean_flow}
\end{figure}

In this example, SPOD is applied to the flow field of the sweeping jet generated from a fluidic oscillator. 
This device is essentially a nozzle with feedback channels, which cause a self-sustained oscillation of the jet.
Figure \ref{fig:fluidix_setup} shows the approximate geometry of this device and indicates the meandering shape of the sweeping jet. 
The shape and motion of the jet resemble a traveling wave.
These devices are used for active flow control applications, where the sweeping motion of the jet allows a much wider actuator spacing resulting in less energy consumption \citep{Woszidlo.2014}. 
The data presented here stem from an experimental setup investigating the spreading and entrainment of sweeping jets \citep{Woszidlo.2015,Ostermann.2015}. 
The data are recorded at a Reynolds number of 37 000 based on the nozzle diameter $D$ and the mean velocity in the nozzle. 
These scales are also used for later calculation of the Strouhal number. 
The mean velocity in figure \ref{fig:fluidix_mean_flow}(a) show that the PIV domain is moved off the jet center towards the lower part of the jet. 
Data points closer than $x/D=2$ were distorted due to laser light reflections. 
The spectral content averaged over the PIV domain (figure \ref{fig:fluidix_mean_flow}(b)) shows a narrow dominant peak and at least five higher harmonics. 
The narrow peaks indicate a stable operation at the fundamental frequency, while the additional peaks suggest more complex dynamics. 
The key challenge of this data set is to accurately reconstruct the sweeping jet dynamics from a truncated measurement domain. 

\begin{figure}
	\begin{flushleft}
	(a)
	\end{flushleft}
	\vspace{3pt}
	\centering
			\includegraphics{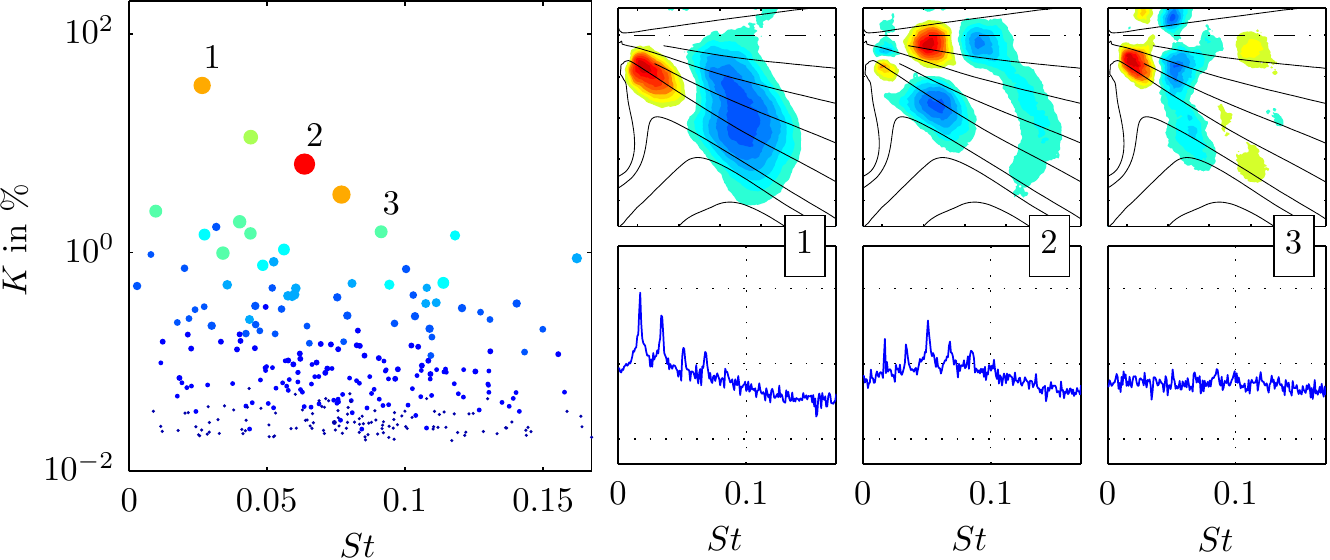}
	\begin{flushleft}
	(b)
	\end{flushleft}
	\vspace{3pt}
	\centering
			\includegraphics{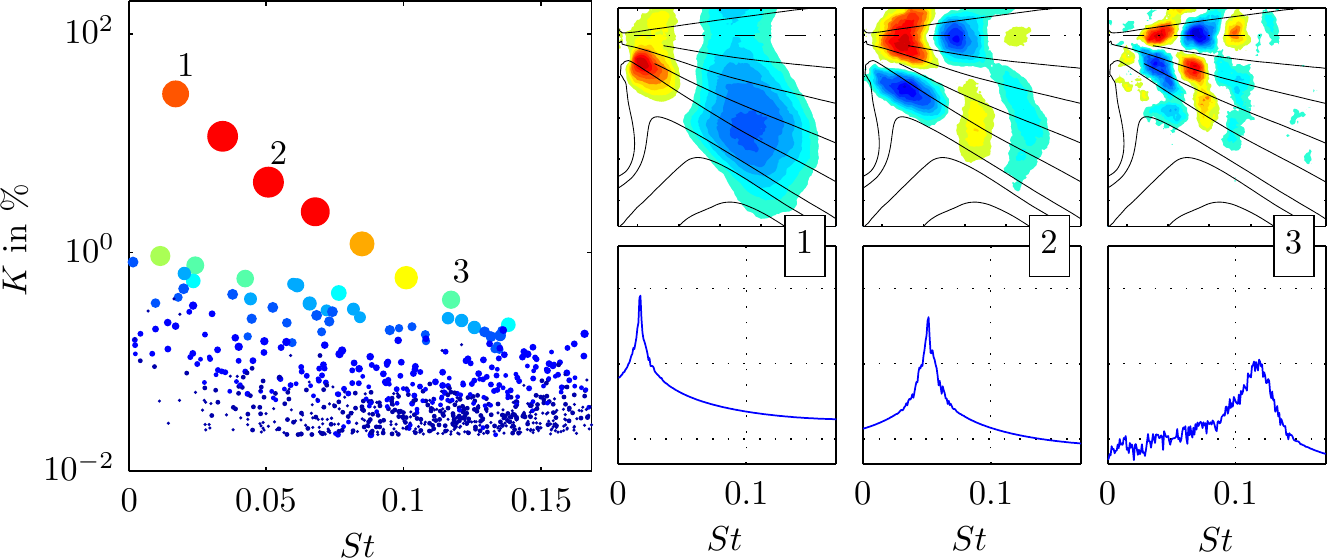}
	\begin{flushleft}
	(c)
	\end{flushleft}
	\vspace{3pt}
	\centering
			\includegraphics{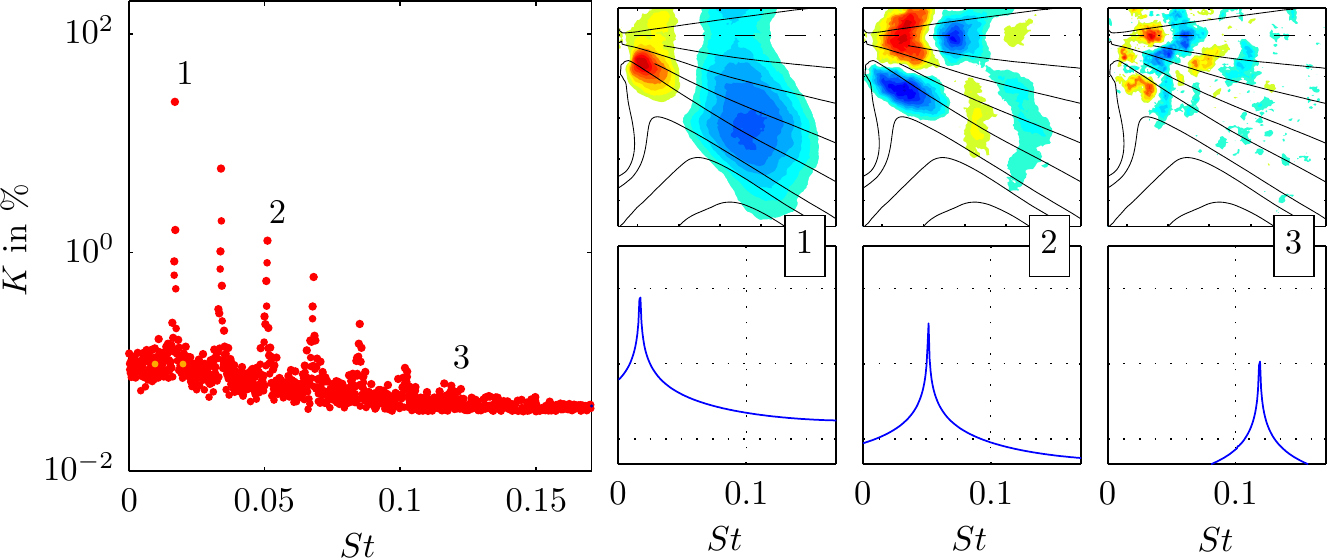}
	\caption{Fluidic oscillator: Results from SPOD for different filter lengths (a) $N_f= 0$ (POD), (b) $ N_f = 30$ (SPOD), and (c) $N_f=2000$ (DFT). For every filter length the SPOD spectrum is displayed as scatter plot (left), where a single dot indicates one mode pair (size and color $C_{i,j}$ in \eqref{eqn:mode_link}). For three selected pairs the spatial modes (upper row) and PSD of the temporal coefficient (lower row) are depicted. They are indicated by numbers in the SPOD spectrum, as well as between the small mode plots. The centerline of is indicated by a dash dotted line.}
	\label{fig:fluidix_spectrum}
\end{figure}

Figure \ref{fig:fluidix_spectrum} shows the results from the SPOD for filter lengths $N_f=0$, 30, and 2000. 
As in the foregoing examples, these filter setting span the range between both limiting cases (from the POD to the DFT). 
The spectrum attained with the POD (figure \ref{fig:fluidix_spectrum}(a)) reveals distinct modes at the fundamental frequency of the oscillator (labeled as 1) and at higher harmonics. 
The mode at the third harmonic frequency (labeled as 2) shows a surprisingly high harmonic correlation. 
The PSD of the mode coefficients reveal that each mode is not limited to a single frequency. 
The additional peaks in the PSD are partly attributed to the fact that only part of the jet is measured. 
During one oscillation period, the jet leaves and enters the measurement domain, creating sharp transitions in the time domain and thus a series of higher harmonics in the frequency domain. 
Due to the purely statistical POD approach, these higher harmonics appear in every mode coefficient, which contradicts the idea of a proper modal decomposition.
The mode that seems to represent the fifth harmonic in the SPOD spectrum (labeled as 3) shows no distinct peak at all in the PSD of the coefficient. 
Thus, the POD of this data set does not provide a proper separation of the fundamental and higher harmonic contributions.

If the SPOD is applied instead, the fundamental and harmonic modes line up perfectly (figure \ref{fig:fluidix_spectrum}(b)). 
Now, the harmonics are separated clearly up to the seventh harmonic. 
The spectral content and spatial shape are further examined for the fundamental, the third and seventh harmonic. 
The PSDs of the mode coefficients reveal narrow spectral bands.
The corresponding mode shapes show an appropriate spatial symmetry, although the PIV domain is cropped shortly above the symmetry line.
It is worth mentioning that the broad peak in the PSD of the seventh harmonic indicates considerable frequency jitter, while the mode shape remains remarkably smooth and symmetric. 

The results obtained with the DFT are presented in figure \ref{fig:fluidix_spectrum}(c). 
The peaks in the SPOD spectrum clearly indicate the fundamental and the first five higher harmonics. 
Their spatial shapes agree well with the SPOD modes, which is not surprising as these modes have narrow spectral bands. 
Note however that each peak is split into several DFT modes, which indicates slight frequency variation. 
This becomes crucial for the higher harmonics, where the frequency jitter is significant and the mode energy is low. 
For the seventh harmonic, the DFT fails to reproduce the structure seen in the SPOD (figure \ref{fig:fluidix_spectrum}(b)). 
The frequency variations detected by the SPOD are simply to high and the mono-frequent energy content too low. This emphasizes the superior noise rejection of the SPOD. 

In this example, the SPOD is superior to the POD and the DFT. 
The energy-ranked POD modes primarily suffer from the incompleteness of the data set. 
This is of immense importance as the relevant domain size is rarely known prior to a set of experiments, or POD analysis. 
The frequency-sharp DFT is insensitive to the domain size, but it fails to reconstruct weak modes with substantial noise and frequency jitter. 
The soft frequency constraint of the SPOD filter operation combines the advantages of both methods and generates a clear mode space. 
The SPOD generates modes with distinct frequencies and mode shapes for modes even weaker than the overall noise level.

\begin{figure}
	\centering
			\includegraphics{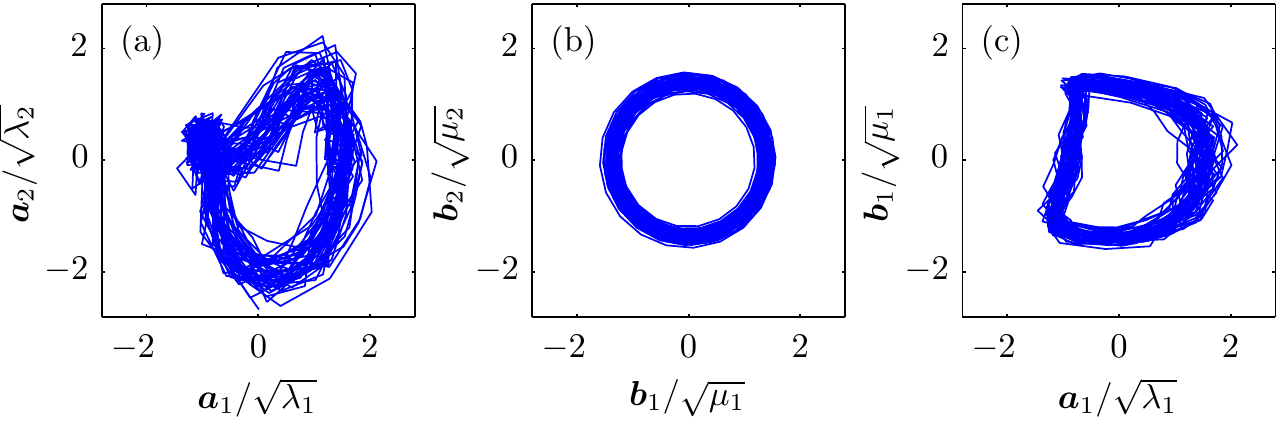}
	\caption{Phase portraits of first temporal coefficients from (a) POD, (b) SPOD, and (c) of both methods against each other.}
	\label{fig:fluidix_SPOD_vs_POD}
\end{figure}

For the fluidic oscillator, the DFT modes are nearly as accurate as those derived with the SPOD. 
The advantages of the SPOD are more obvious for less mono-frequent flow dynamics (see the previous examples). 
However, an additional advantage of the SPOD against the DFT is that it provides a reliable estimate of the oscillatory phase by accounting for the frequency jitter. 
Similar approaches, which also produce satisfactory results for the current case are described by \citet{Ostermann.2014}, but again, the scope of SPOD is beyond this particular application.
Figure \ref{fig:fluidix_SPOD_vs_POD}(a,b) shows the phase portraits (Lissajous figures) of the temporal coefficients of the two most energetic POD ($\av_i$) and SPOD ($\bv_i$) modes, respectively. 
The trajectory of the POD coefficients does not follow a clear circle that would indicate the limit-cycle. 
It rather follows a third of a circle and then collapses at one point. 
The coefficient of the SPOD modes follows a clear circle and the instantaneous phase and instantaneous frequency can consistently be deduced. 
A comparison of the first mode coefficient from both methods is shown in figure \ref{fig:fluidix_SPOD_vs_POD}(c). 
It reveals that half of the period is cut out for the POD ($\av_1$). 
This corresponds to the sweeping jet leaving the measurement domain, where the energy-based POD ``sees'' no jet.  
The SPOD properly recovers the temporal dynamics over the entire oscillation period. 
Furthermore, note that the SPOD produces coefficients with smooth temporal dynamics, while the POD coefficients show rather erratic movements.
This is particularly important for reduced order modeling \citep{Luchtenburg.2009}, phase averaging, and extended POD \citep{Boree.2003}. 
Practically, most of the further processing is eased if there is less noise.

\section{Summary and conclusion}\label{sec:summary_conclusion}

\subsection{Properties and capabilities of the proposed method}\label{summary}

The SPOD is introduced as an extension of the POD for time-resolved data. 
This novel method involves a filter operation on the diagonal elements of the snapshot correlation matrix. 
The procedure is closely related to the classic snapshot POD with a negligible increase of algorithmic complexity and numerical costs.  
The SPOD filter allows for a continuous fading from the energetic optimality of POD to the spectral purity of DFT. 
It is conceptualized in a general form, with the POD and the DFT as the limiting cases. 
The concept of SPOD was developed trough our experience with experimental data processing, and not from a constraint optimization problem.
It arose from the desperate need for a method that applies to a wide range of turbulent flows at minimum user input.  
The SPOD is motivated based on theoretical considerations, where it is interpreted as a short time linearization of the flow dynamics. 

The key feature of the SPOD is the smoothing of the diagonal elements of the correlation matrix.
This filter operation is shown to constrain the growth rates, amplitudes and frequencies of the SPOD modes.
By setting the filter width, one gains control over the spectral bandwidth of the single modes. 
When the filter is set to the maximum length, the modes are assumed to be strictly periodic and the SPOD converges to the DFT. 

\begin{figure}
	\centering
			\includegraphics{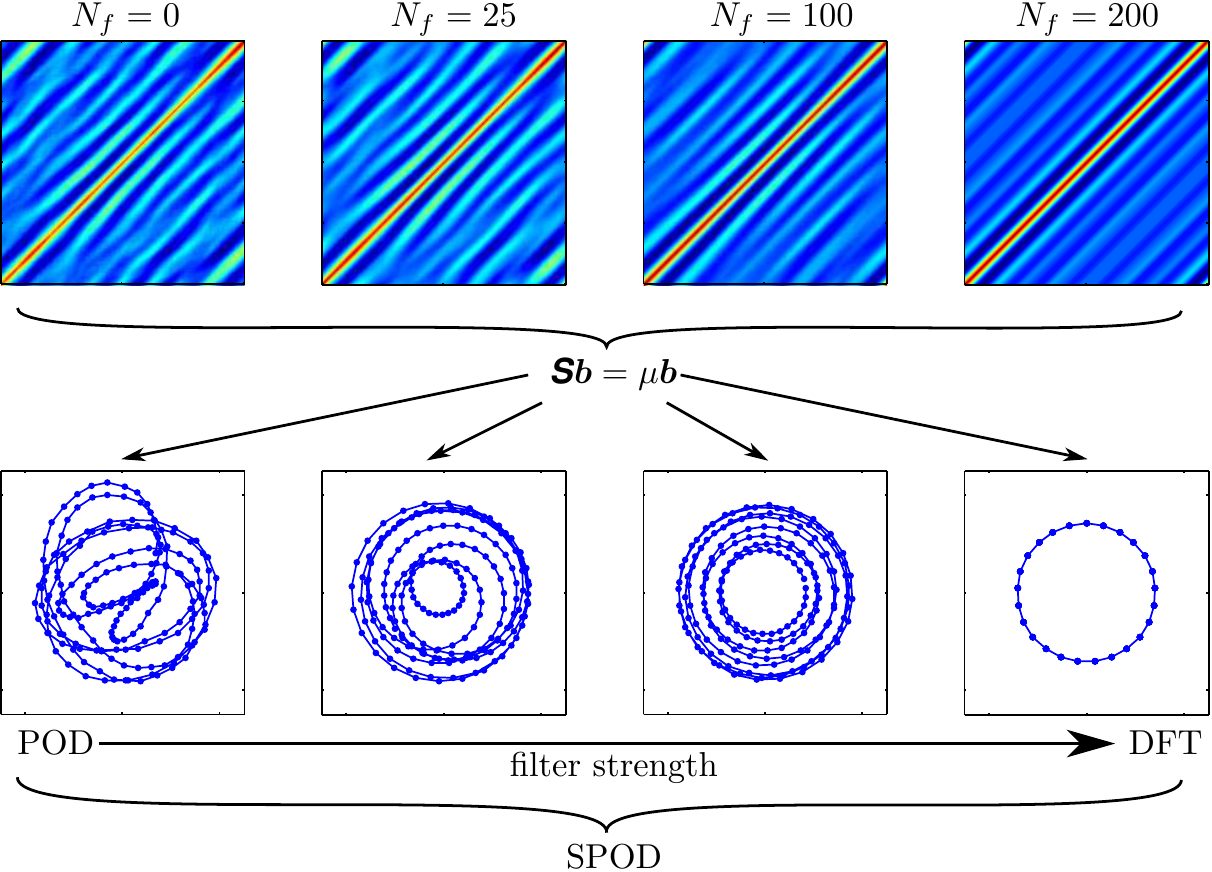}
	\caption{ Schematic describing the main properties of the SPOD for increasing filter strength (from left to right). The top row shows pseudo-color plots of the filtered correlation matrix matrix ($\Smat$). The phase portraits of the corresponding first two modes ($\bv_1$ and $\bv_2$) that describe the dominant oscillations are shown below. The axes of the plots shown here are the same as for the plots in figure \ref{fig:corr_mat} and \ref{fig:fluidix_SPOD_vs_POD}, respectively. The graphs are based on the data already presented in section \ref{corr_matrix} and the SPOD is calculated from 200 snapshots.}
	\label{fig:SPDO_flowmap}
\end{figure}

The principle of the SPOD is graphically summarized in figure \ref{fig:SPDO_flowmap}. 
The images in the first row show the filtered correlation matrix at different filter widths $N_f$.
The images below depict the phase portraits of the leading two modes (compare figure \ref{fig:fluidix_SPOD_vs_POD}).
It is apparent that the increased diagonal similarity of the correlation matrix, that goes in hand with the increased filter width, successively limits the temporal variations of the mode amplitude and frequency until a stable limit cycle is reached.  

The application of SPOD to the flow field of a swirl-stabilized combustor, an airfoil with Gurney flap and a fluidic oscillator revealed different advantageous features of the SPOD in comparison to POD or DFT. 
For every single case there exist other suitable methods, which may perform equally well as the SPOD, but none of them is as versatile as the proposed method. 

The main advantages can be summarized as follows:
\begin{description}
	\item[Separation of structures: ] The soft spectral constraint of the SPOD allows for a much better separation of individual fluid dynamic phenomena into single modes, whereas POD or DFT mix or spread them among several modes.
	\item[Noise rejection: ] SPOD it is insensitive to noise and even recovers dynamics that are weaker than the overall noise level.
	\item[Data completion: ] SPOD can eliminate degradation of temporal dynamics of partially recorded phenomena. 
	\item[Plain dynamics: ] The mode coefficients are smooth in time and they feature adjustable variations of frequency and amplitude (set by the filter size).
\end{description}

The characteristics of the SPOD modes ease further processing such as the identification of linked modes, comparisons against other simultaneously acquired measurements (phase averaging or extended POD) and the identification of reduced order models. The SPOD may also provide a better basis for modal representation of snapshots as input for a DMD, as pursued in section \ref{SPOD_mode_link}.

\subsection{Concluding remarks}\label{conclusion}
The SPOD has proven to be a reliable method to extract distinct phenomena from turbulent flows. 
It was not derived from purely mathematical considerations, but rather evolved from practical data processing. 
Nonetheless, the method has well defined upper and lower bounds and generates modes that can be easily interpreted. 
As shown in the considered examples, SPOD is a simple way to extract coherent structures from turbulent flows, where the POD or the DFT failed to provide valuable results. 
The SPOD constrains the spectral content, but retains the modal sparsity of the POD. 

There are certainly plenty of other cases where this new method will ease the identification of hidden coherent structures.
Its true benefit lies in the fact that only one assumption is made about the investigated flow dynamics, which is the filter timescale. 
This can also be understood as an inertia imposed on the mode dynamics, limiting the rate of change of the frequency and amplitude. 
The choice of this timescale can be assessed from the flow's dominant frequency or convective timescale, as shown in this article. 
The authors hope that the SPOD will give access to new fluid dynamical phenomena and enriches the available methods.

\subsection{Acknowledgment}\label{acknowledgment}
The authors kindly acknowledge the stimulating discussions with Lothar Rukes and the generous provision of experimental data from Alena Bach and Florian Ostermann. 
The funding from the German Research Foundation (DFG Project PA 920/30-1) is also gratefully acknowledged. 

\begin{appendix}

\section{The spatial correlation version of SPOD}\label{spatial_SPOD}

The original POD can either be calculated from a spatial or temporal correlation, which allows a computationally efficient calculation by restricting the size of the problem to the number of the snapshots or the number of grid points. 
Similarly, the SPOD has a spatial correlation counterpart, which is computationally more efficient if the number of snapshots is much larger than the number of grid points. 
This approach is slightly more complex and less intuitive than the snapshot version. 
Nevertheless, it is very valuable if long time series of few sensors are supposed to be decomposed into proper modes to perform an extended POD or to derive the phase of an oscillatory mode from the measurements. 
Assume a simultaneous multi point pressure measurement that shall be decomposed with SPOD. This series is decomposed as
\begin{align}
P(x_i,t) = \overline{p}(x_i) + \sum_{s=1}^N b_s(t) \Psi_s(x_i),
\end{align}
where the number of measured positions $M$ is much smaller than the number of samples $N$. 
The number of samples may easily reach a million or more, which complicates the solution in terms of memory requirements for the composition of the temporal correlation matrix and in terms of computational time for the solution of the eigenvalue problem. 
Therefore the temporal correlation described in section \ref{spectral_POD} is not feasible in this case. 
Instead, the spatial correlation should be employed, as outlined in this section. 
The multi time shift correlation tensor for the spatial SPOD reads
\begin{align}
S_{i,j,k,l} = \frac{\sqrt{g_k g_l}}{M N \deltat} \int{p(x_i,t-k\deltat) p(x_j,t-l\deltat)}\mathrm{d}t\\
\ i,j = 1...M\ ; \ k,l = -N_f...N_f ,\nonumber
\end{align}
where $p = P-\overline{p}$ is the fluctuating part of the pressure and $g_k$ are the filter coefficients. 
For numerical implementation this is reshaped to a matrix such that $S_{i,j,k,l} = \widetilde{S}_{(i+k*M),(j+l*M)}$, but for the theoretical description the tensor notation is retained. 
The correlation tensor is decomposed in eigenvalues and eigenvectors, such that
\begin{align}
\sum_{l=-N_f}^{N_f}{\sum_{j=1}^M{S_{i,j,k,l} \widetilde{\Psi}_s(x_j,\tau_l)}} = \mu_s \widetilde{\Psi}_s(x_i,\tau_k)  \ ; \quad \mu_1 \ge \mu_2 \ge \cdots \ge \mu_N \ge 0. \label{eqn:spatial_eigen},
\end{align}
where $\tau_k = k\deltat$.
The eigenvector $\widetilde{\Psi}_s$ constitutes a discrete convolution filter, which is applied to the time series to obtain the mode coefficients
\begin{align}
b_s(t) = \sum_{k=-N_f}^{N_f}{\sum_{i=1}^{M}{\sqrt{\frac{g_k}{M}}\widetilde{\Psi}_s(x_i,\tau_k)\ p(x_i,t-\tau_k)}}.
\end{align}
The spatial mode is the central part ($\tau_k =0$) of the convolution filter $\Psi_s(x_i) = \widetilde{\Psi}_s(x_i,0)$. 
The entire eigenvectors $\widetilde{\Psi}_s$ can be understood as a data driven filter bank, which allows for decomposition of time series into modal contributions.
It might be applied to a single sensor, where each mode represents a certain spectral band of the signal.
The principal approach is comparable to the empirical mode decomposition \citep{Huang.1998}, but the SPOD can also handle multiple sensors.
In case of multiple sensors, it gives excellent result when the phase of a dominant oscillation has to be reconstructed from pressure measurements.
The approach outlined in this section is similar to the multi time delay POD phase estimation pursued by \citet{Hosseini.2015}.

In contrast to the snapshot version, the computational cost of the spatial version of SPOD scales with the filter size. It is only more efficient than the snapshot approach if $M(2N_f+1) < N$.

\section{Properties of the SPOD modes}\label{SPOD_mode_properties}

In section \ref{spectral_POD} it was shortly mentioned that the spatial SPOD modes are no longer orthogonal, which is only part of the truth. 
If the spatial mode $\widetilde{\Psi}$ \eqref{eqn:spatial_eigen} together with all of the temporally shifted instances is considered, they are orthonormal
\begin{align}
\sum_{l=-N_f}^{N_f}{\sum_{k=1}^M{ \frac{g_k}{M} \ \widetilde{\Psi}_i(x_k,\tau_l) \ \widetilde{\Psi}_j(x_k,\tau_l) }} = \delta_{i,j}.
\end{align}
The snapshot based calculation introduced in section \ref{spectral_POD}, however, is restricted to the zero time delay ($\tau =0$) part of the spatial mode. 
Furthermore, the decomposition of the data into modal contributions is only feasible for time independent spatial modes. 
This limitation to one part of the the spatial mode $\Psi_s(x_i) = \widetilde{\Psi}_s(x_i,\tau=0)$ introduces some imperfections. 
The selected modes for the decomposition are neither normal nor orthogonal
\begin{align}
\frac{1}{M} {\sum_{k=1}^M{\Psi_i(x_k) \ \Psi}_j(x_k) } \ne \delta_{i,j}.
\end{align}
The loss of normality is a fact, whereas the norm of the spatial modes gives further insights to the data set. The norm 
\begin{align}
\zeta_i = \sqrt{\frac{1}{M}\sum_{k=1}^M{\Psi_i^2(x_k) }}
\end{align}
indicates how well a single mode is represented by the investigated data set. 

With the application of the filter \eqref{eqn:SPOD_filter}, an idealized correlation matrix is constructed that delivers modes, which are more or less captured by the initial data set. 
This fact is reflected by the deviation of the mode norm $\zeta_i$ from one. 
Consider for example the measurement of the sweeping jet. There, the fundamental mode is only partially captured as shown in figure \ref{fig:fluidix_SPOD_vs_POD}. 
With the SPOD, the missing data are completed and a SPOD mode pair with equal energy levels $\mu_i$ is obtained. 
However, for the construction of the spatial modes the coefficients are projected onto the original data \eqref{eqn:SPOD_temporal_projection}. 
There, the imperfect representation of one of the two modes re-enters the processing. 
For the sweeping jet's leading mode pair the norm $\zeta_i$ of one mode is clearly below the other, but they approximately add up to one. 
Therefore, the eigenvalues $\mu_i$ describe the idealized energy content of the single modes and the norm of the spatial mode $\zeta_i$ corrects the deficits in comparison to the actual data set.
The limiting SPOD cases (POD and DFT) do not show this deficits. 
The POD modes are already normalized $\zeta_i=1$ and for the DFT, the modes pair perfectly, while the norm of these pairs ($i,j$) exactly add up to one ($\zeta_i + \zeta_j = 1$).

\end{appendix}

\bibliographystyle{jfm}
\bibliography{references}

\end{document}